\documentclass{PoS}
\usepackage{cite}
\usepackage{amsmath}

\title{Progress on soft gluon exponentiation and long-distance singularities}
\ShortTitle{Progress on soft gluon exponentiation and long-distance singularities}

\author{\speaker{Einan Gardi}
       \\
       Higgs Centre for Theoretical Physics, School of Physics and Astronomy, \\
The University of Edinburgh,  
Edinburgh EH9 3JZ, Scotland, UK\\
       E-mail: \email{Einan.Gardi@ed.ac.uk}}

\abstract{I review the recent progress in studying long-distance singularities in gauge-theory scattering amplitudes in terms of Wilson lines. The non-Abelian exponentiation theorem, which has been recently generalised to the case of multi-leg amplitudes, states that diagrams exponentiate such that the colour factors in the exponent are fully connected. After a brief review of the diagrammatic approach to soft gluon exponentiation, I sketch the method we used to prove the theorem and illustrate how connected colour factors emerge in the exponent in webs that are formed by sets of multiple-gluon-exchange diagrams. In the second part of the talk I report on recent progress in evaluating the corresponding integrals, where a major simplification is achieved upon formulating the calculation in terms of subtracted webs. I argue that the contributions of all multiple-gluon-exchange diagrams to the soft anomalous dimension take the form of products of specific polylogarithmic functions, each depending on a single cusp angle.}

\FullConference{11th International Symposium on Radiative Corrections (Applications of Quantum Field Theory to Phenomenology) \\
		 22-27 September 2013\\
		 Lumley Castle Hotel, Durham, UK}

\begin{document}

\section{Introduction}

I will discuss soft-gluon exponentiation and the recent progress towards determining soft singularities in multi-leg scattering amplitudes at three loops.
The motivation of this research programme is to extend our knowledge of long-distance singularities, which is a key to precision cross-section calculations. Progress on this front will feed into creating effective subtraction methods for combining real and virtual corrections, and lead to more precise resummation of logarithms in a broad range of observables.
Beyond these applications there is also a strong theoretical motivation to further our understanding of the mathematical structure of scattering amplitudes in general. Here infrared singularities have a special role because their structure is vastly simpler than the finite parts of the amplitudes, so real progress towards multi-loop and even all-order results can indeed be made. It should also be noted that the long-distance singularity structure is rather similar for all non-Abelian gauge theories. Specifically the matter contents of the theory will not play any important role in our discussion.

In this talk I will be specifically interested in multi-leg processes, and I will not make any assumptions about their spins or their colours - in particular no large-$N_c$ approximation will be taken. The strategy I take here in studying the infrared limit of scattering amplitudes is to compute the renormalization of a product of semi-infinite Wilson lines, all meeting at a point,
\begin{equation}
\label{S}
S=\exp\left[ w \right]= \left<\Phi_{\beta_1}\,\otimes\,\Phi_{\beta_2}\,\otimes\,\ldots \Phi_{\beta_L}\right>\,.
\end{equation}
Each Wilson line is defined by 
\begin{equation}
\label{path-ordered_exp}
\Phi^{(l)}_{\beta_l}\,\equiv\,
{\cal P}\exp\left[{\mathrm i}g_s \int_0^{\infty}dt \beta_l\cdot {A}(t\beta_l)\right]\,,
\end{equation}
where we suppressed the colour indices and explicitly displayed the dependence on the 4-velocity~$\beta_l$. Here ${\cal P}$ denotes path-ordering of the colour generators along the Wilson line.
The $L$ Wilson lines originate in the eikonal approximation to the corresponding $L$ highly energetic coloured particles participating in the scattering process; these dictate the directions and colour charges of the lines. In order to avoid collinear singularities (and preserve multiplicative renormalizability, see below) we will use massive, or non-lightlike Wilson lines, where the 4-velocities obey $\beta_l^2\neq 0$. When considering massless partons we will be able to take the limit $\beta_l^2\to 0$ at the end and combine the result with the relevant jet functions (see e.g. Refs.~\cite{Aybat:2006wq,Aybat:2006mz,Gardi:2009zv,Ferroglia:2009ep}). $S$ with timelike Wilson lines, $\beta_l^2>0$, is directly relevant for heavy quarks.
\begin{figure} [htb]
\begin{center}
\includegraphics[width=.25\textwidth]{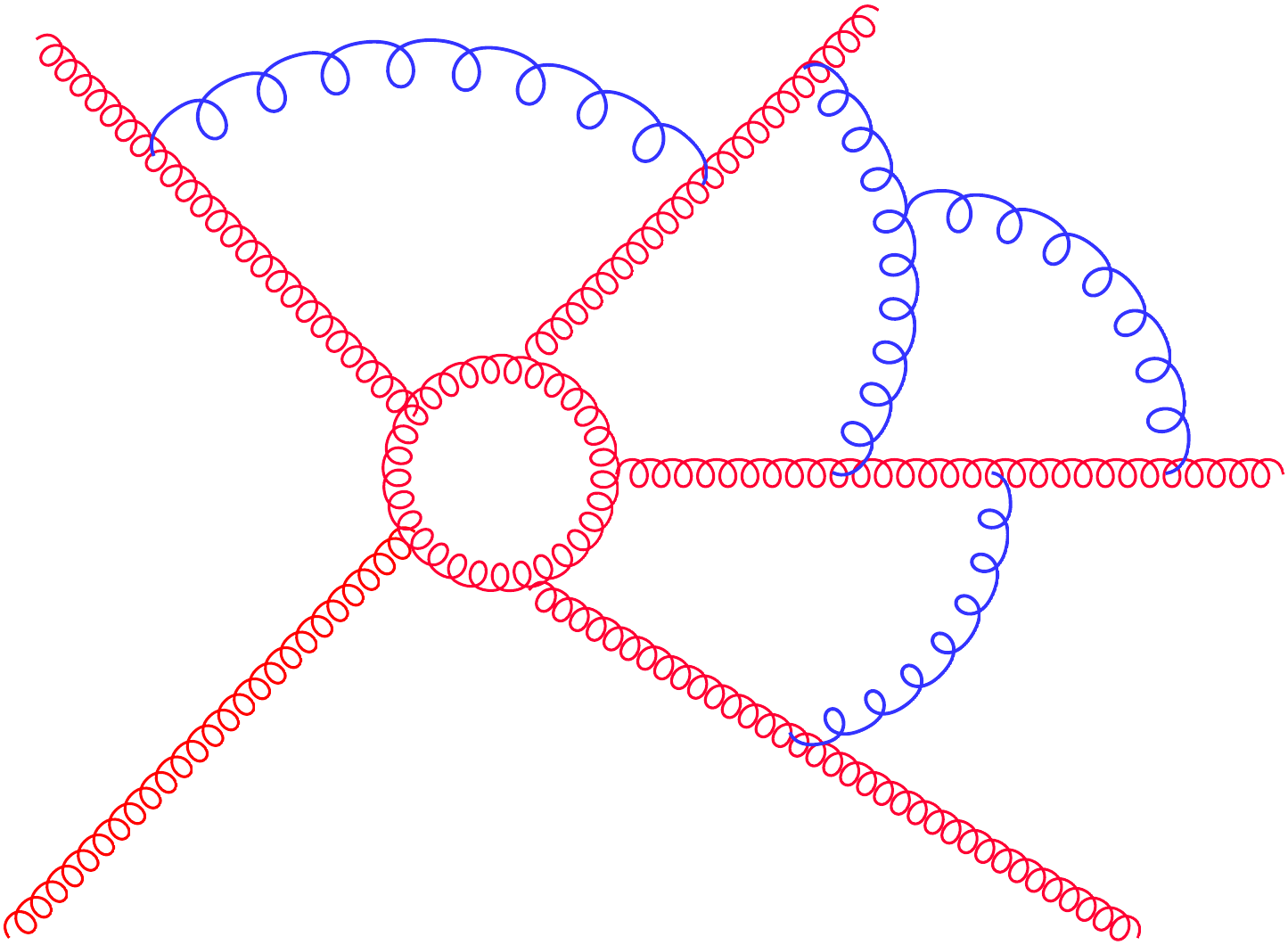}
\includegraphics[width=.125\textwidth]{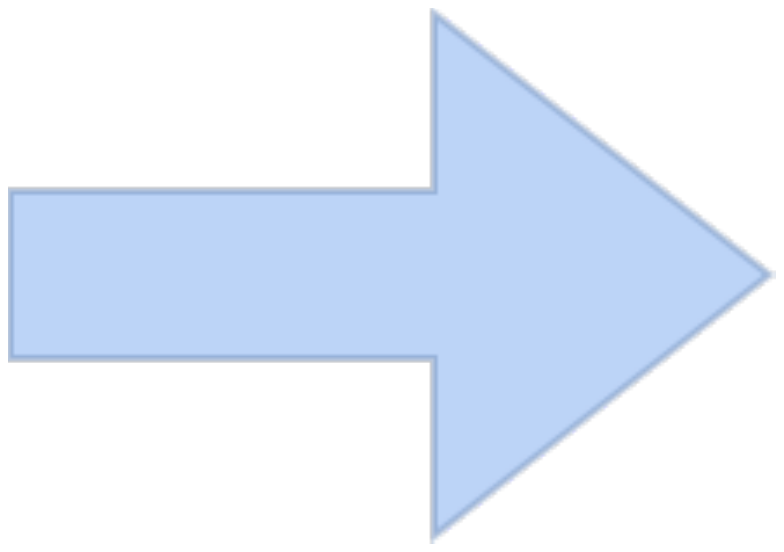}
\includegraphics[width=.25\textwidth]{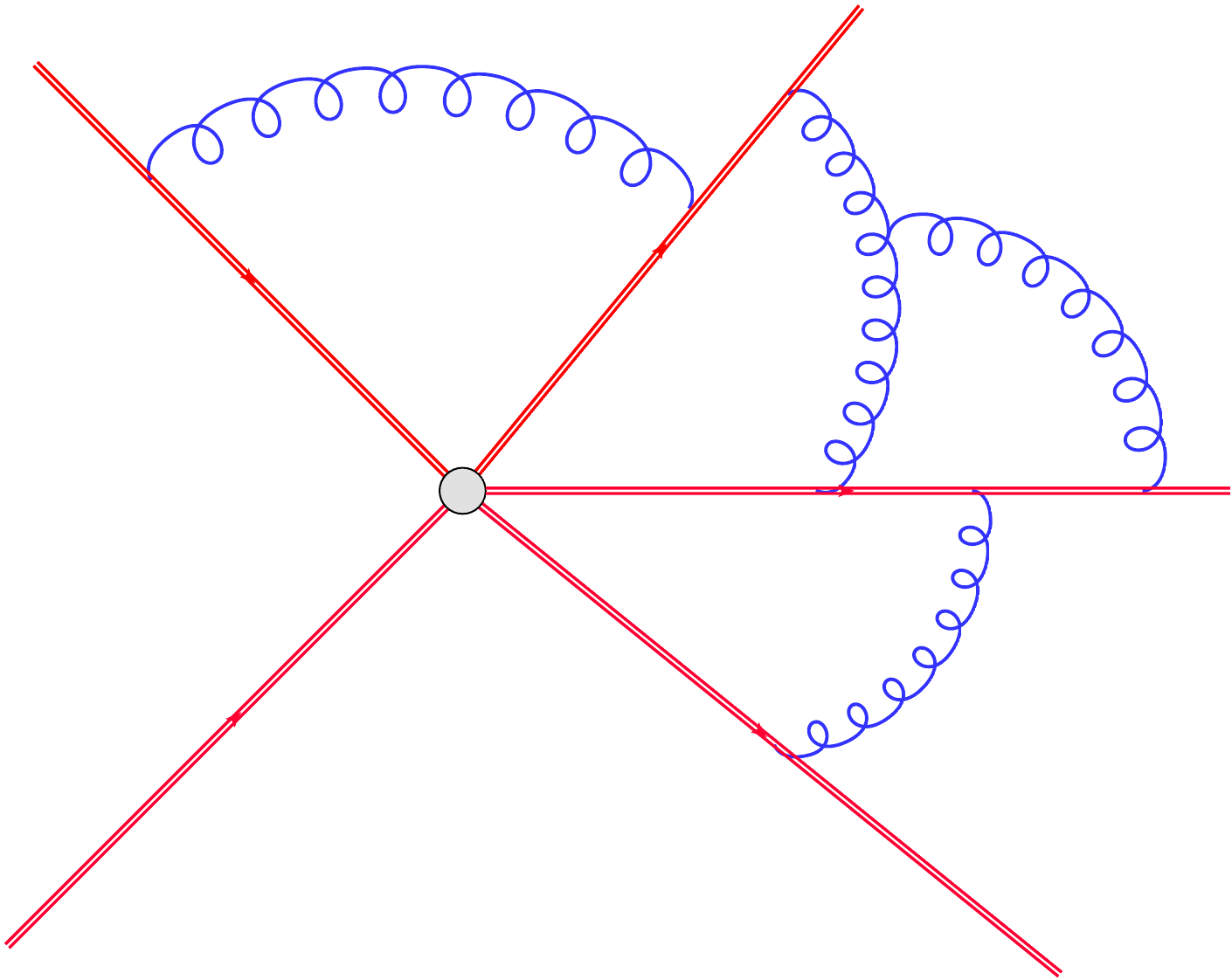} 
\end{center}
\caption{A Feynman diagram contributing to an amplitude involving five hard external partons and the corresponding product of five Wilson lines extending from the hard interaction point, where they all meet, to infinity. The soft gluons are the same in both.} \label{fig1}
\end{figure}

In dimensional regularization $S$ presents a rather remarkable relation between the ultraviolet and the infrared singularity structure~\cite{Korchemsky:1987wg} owing to the fact that scaleless integrals vanish identically. Instead of computing infrared singularities we will compute the renormalization of the vertex formed by the product of Wilson lines in eq.~(\ref{S}). This operator renormalizes multiplicatively~\cite{Polyakov:1980ca,Arefeva:1980zd,Dotsenko:1979wb,Brandt:1981kf}:
\begin{equation}
S_{\rm ren.}(\epsilon_{\rm IR},\mu)=S_{{\rm UV}+{\rm IR}}\, Z(\epsilon_{\rm UV},\mu).
\end{equation}
 In the absence of any cutoff all radiative corrections vanish and $S_{{\rm UV}+{\rm IR}}=1$, which implies 
\begin{equation}
\label{IR_UV}
S_{\rm ren.}(\epsilon_{\rm IR},\mu)= Z(\epsilon_{\rm UV},\mu).
\end{equation}
The logarithmic derivative of the $Z$ factor is the so-called \emph{soft anomalous dimension}: ${dZ}/{d\ln \mu}=-Z\,\Gamma$. The anomalous dimension $\Gamma$ is finite and it carries all the necessary information concerning soft singularities. $Z$ can be solved for and be expressed as an ordered exponential of an integral of $\Gamma$ over the scale (in a more explicit form it is given by eq.~(\ref{Z_exp}) below). We thus see that exponentiation largely dictates the structure of long-distance singularities of amplitudes, making it a very useful property, see e.g. Refs.~\cite{Catani:1998bh,Sterman:2002qn}.

The remainder of the talk consists of two parts. The first is a brief introduction to the diagrammatic picture of non-Abelian exponentiation. The second deals with calculations of a class of diagrams contributing to $\Gamma$, which we now understand well, namely those involving gluon exchanges between the Wilson lines, without any three or four gluon vertices. This calculation is greatly simplified  thanks to the workings of non-Abelian exponentiation.

\section{The non-Abelian exponentiation theorem\label{non-abelian_exp_theorem}}
\begin{figure}[htb]
\begin{center}
\vspace*{-140pt}
\includegraphics[width=.6\textwidth]{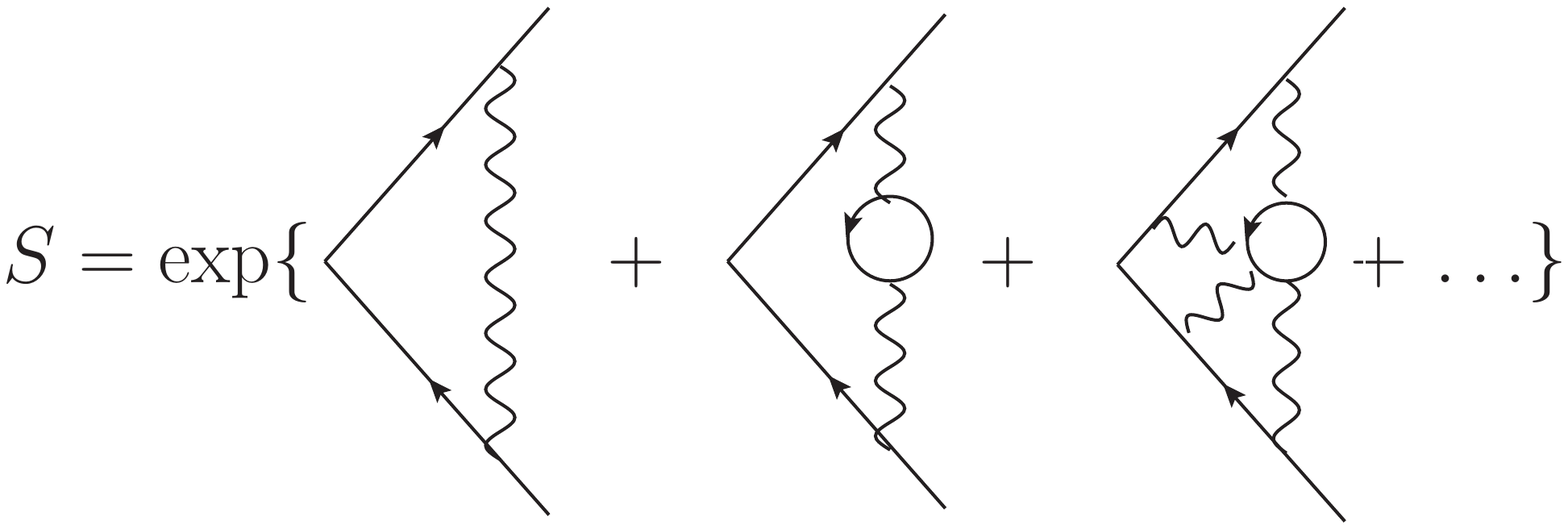}
\vskip -120pt
\hspace{40pt}
\includegraphics[width=.5\textwidth]{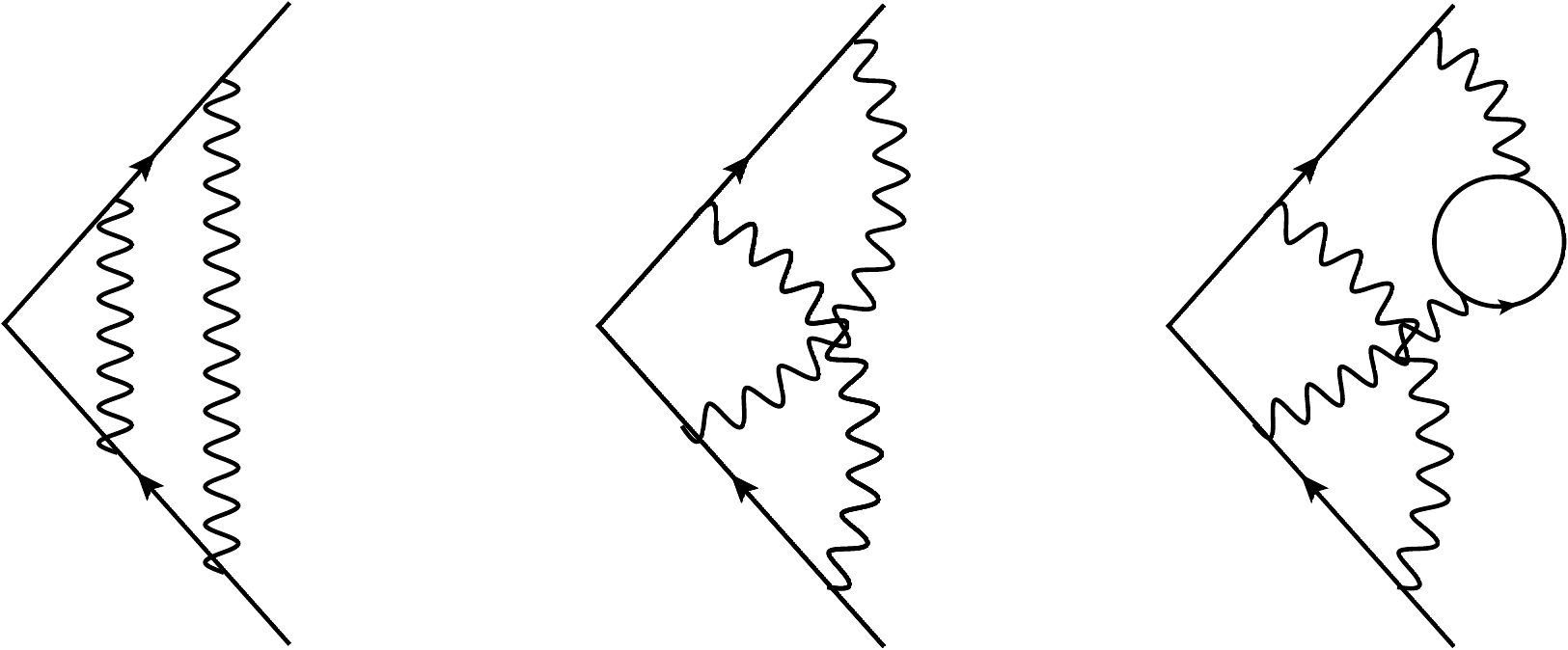}
\caption{Exponentiation for a product of two Wilson-line rays in the Abelian theory. Top: the first few connected graphs which build up the exponent. Bottom: examples of non-connected diagrams that are generated by expanding the exponential.} \label{Abelian_exp}
\end{center}
\end{figure}

Let us now discuss soft gluon exponentiation. The starting point is the observation that when considering a Wilson-line correlator, as in eq.~(\ref{S}), the exponent $w$ takes a simple form. Consequently it is useful to compute it directly.
This was first understood in the context of the Abelian theory in the 1960's~\cite{Yennie:1961ad}; in this case the exponent $w$ only receives contributions from connected graphs (throughout our discussion `connected' should be understood as referring to the graph after removing the Wilson lines). All non-connected diagrams are reproduced upon expanding the exponential. This is illustrated in figure~\ref{Abelian_exp}.

The next step was taken in the 1980's~\cite{Sterman:1981jc,Gatheral:1983cz,Frenkel:1984pz}, when the non-Abelian exponentiation theorem was first formulated. This was done in the context of a Wilson loop, or two Wilson lines, corresponding to a colour singlet form factor (for a review see~\cite{Berger:2003zh}).  
The generalization to a product of more than two Wilson lines, as relevant for QCD hard scattering, was only made over the last three years~\cite{Gardi:2010rn,Mitov:2010rp,Gardi:2011wa,Gardi:2011yz,Dukes:2013wa,Dukes:2013gea,Gardi:2013ita}.  

\subsection{Exponentiation for two Wilson lines}

\begin{figure}[htb]
\begin{center}
\includegraphics[width=.6\textwidth]{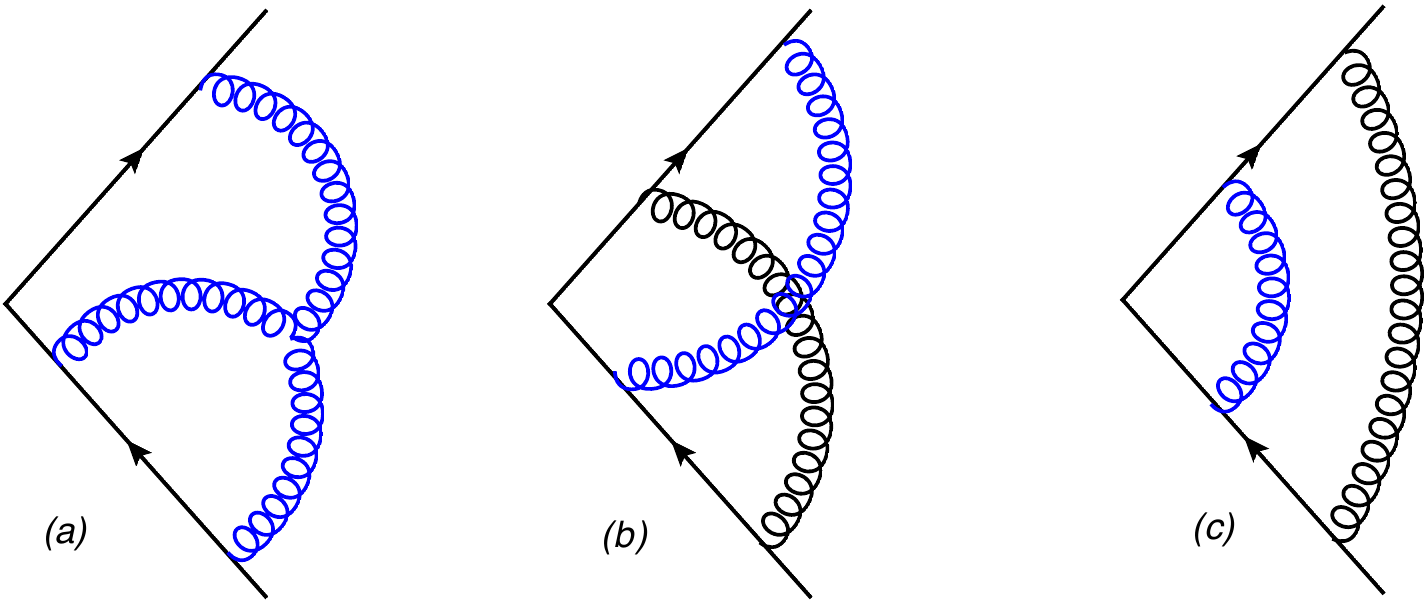}
\caption{Examples of reducible (c), and irreducible (a and b), two-loop diagrams in the non-Abelian two-line case. In (c) the inner gluon can be shrunk to the cusp without affecting the outer one, resulting in a leading $1/\epsilon^2$ ultraviolet singularity; this may be contrasted with diagrams (a) and (b) which have a single ultraviolet pole. This illustrates the connection between the irreducible colour structure and the absence of subdivergences in the renormalization of the cusp. } \label{non-Abelian}
\end{center}
\end{figure}
Let me first describe the part of the picture that was known until 2010, namely considering two Wilson lines. Here there is a simple topological criterion for determining which diagrams contribute to the exponent and which do not. To make this separation one defines a \emph{reducible diagram} as one whose colour factor can be written as the product of the colour factors of its subdiagrams (an example is provided in figure~\ref{non-Abelian}). In this language the non-Abelian exponentiation theorem of~\cite{Sterman:1981jc,Gatheral:1983cz,Frenkel:1984pz} states that the exponent only receives contributions from irreducible diagrams, while the reducible ones are fully reproduced by exponentiation of lower-order diagrams corresponding to their subdiagrams.
Furthermore, the colour factors of the diagrams that enter the exponent (ECF, Exponentiated Colour Factors) are different from the ordinary colour factors of these diagrams, and are given by the ``connected'' or ``non-Abelian'' parts of the latter. For example, the ECF of diagram (b) in figure~\ref{non-Abelian} is\footnote{Here $C_R$ is the quadratic Casimir in the representation $R$ where $C_A$ corresponds to the Adjoint representation. $R$ is the representation of the two Wilson lines in figure~\ref{non-Abelian}.} $-\frac12 C_R C_A$, which is the non-Abelian part of its ordinary colour factor, $C_R(C_R-\frac12C_A)$.

Beyond the colour structure aspect, reducible diagrams also have the property that they have subdivergences associated with the cusp (this is illustrated in the example of figure~\ref{non-Abelian}); such subdivergences never appear in irreducible diagrams. 
The absence of subdivergences in diagrams that contribute to the exponent is essential for the multiplicatively renormalizability of the cusp~\cite{Polyakov:1980ca,Arefeva:1980zd,Dotsenko:1979wb,Brandt:1981kf}. This is related to the fact that upon solving the renormalization group equation for $Z$ in terms of $\Gamma$ for the two-line case one obtains an integral over the scale, which produces\footnote{Higher poles in $\epsilon$ are generated by running coupling effects, but these are distinct from renormalization of the cusp.} \emph{a single} $1/\epsilon$ pole. The connection between the irreducibility of the colour factor and the absence of subdivergences in the exponent, which in turn relates to the renormalization properties of the operator, will be central to understanding the generalization to the multi-leg case.

\subsection{Exponentiation in the multi-line case and web mixing matrices}

\begin{figure}[htb]
\begin{center}
\includegraphics[width=.35\textwidth]{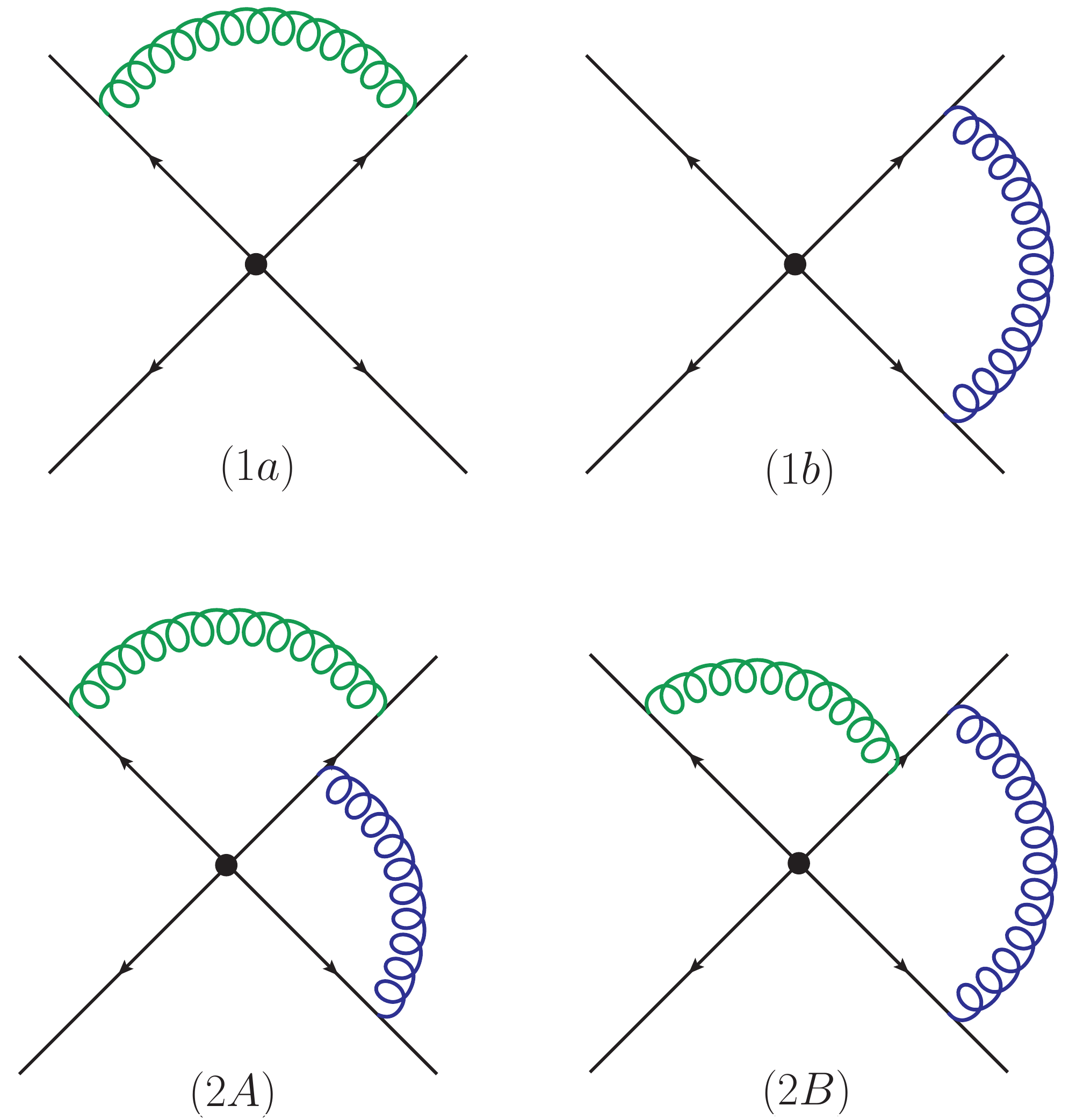}
\caption{One and two-loop gluon-exchange diagrams contributing to the renormalization of the product of three or more Wilson lines. Diagrams (1a) and (1b) correspond to the subdiagrams of (2A) and (2B); clearly the latter are reducible diagrams. Nevertheless, as shown in the text these two-loop diagrams are not fully reproduced by exponentiation of the one-loop ones. Rather, one must include the antisymmetric combination of the two colour factors times the antisymmetric combination of the two kinematic integrals in the exponent. } \label{4legs_two_loops}
\end{center}
\end{figure}

Upon considering several Wilson lines ones quickly realises that the non-Abelian exponentiation picture described above breaks down: here reducible diagrams do contribute to the exponent. To see this consider the example of figure~\ref{4legs_two_loops}. 
Let us write the contribution of each diagram $D$ as an explicit colour factor $C_D$ times a kinematic-dependent integral ${\cal F}_D$. The two one-loop diagrams of figure~\ref{4legs_two_loops} yield:
\begin{align}
\begin{split}
D_{(1a)}+D_{(1b)}={\cal F}_{(1a)} T_1\cdot T_2 +{\cal F}_{(1b)}T_2\cdot T_3\,,
\end{split}
\end{align}
where $T_i$ is a colour generator on line $i$, and $T_1\cdot T_2=T_1^{(a)} T_2^{(a)}$, under summation convention over $a$. The result of exponentiating these one-loop graphs, expanded to two-loop order, yields:
\begin{align}
\begin{split}
\frac12\left[D_{(1a)}+D_{(1b)}\right]^2&=
\frac12\left[ D_{(1a)}D_{(1b)}+D_{(1b)}D_{(1a)}+\cdots\right]\\
&=\frac12{\cal F}_{(1a)}{\cal F}_{(1b)} \,\,{\color[rgb]{1,0.0,0.0}{\mathrm T}_1^{(a)}\left[{\mathrm T}_2^{(a)}{\mathrm T}_2^{(b)}+{\mathrm T}_2^{(b)}{\mathrm T}_2^{(a)}\right]{\mathrm T}_3^{(b)} }+\cdots\\
&=\frac12\Big[{\cal F}_{(2a)}+{\cal F}_{(2b)}\Big]\,\,\,\,{\color[rgb]{1,0.0,0.0}\Big[C_{(2a)}+C_{(2b)}\Big]}+\cdots
\end{split}
\end{align}
which is different from the sum of the two two-loop diagrams:
\begin{align}
\begin{split}
D_{(2a)}+D_{2b}={\cal F}_{(2a)} {\color[rgb]{1,0.0,0.0}C_{(2a)} }+{\cal F}_{(2b)}{\color[rgb]{1,0.0,0.0}C_{(2b)}}\,.
\end{split}
\end{align}
We see that only the symmetric combination of colour factors of the diagrams is reproduced by expanding the exponential, implying that we need to incorporate in the exponent the anti-symmetric contribution,
\begin{align}
\label{two_loop_example_exp}
\begin{split}
\frac12 \underbrace{
\Big[{\cal F}_{(2a)}-{\cal F}_{(2b)}\Big]}_{{\cal O}(1/\epsilon)}\,\,\,\,{\color[rgb]{1,0.0,0.0}
\underbrace{\Big[C_{(2a)}-C_{(2b)}\Big]}_{{\rm i}f^{abc} {\rm T}_1^{(a)} {\rm T}_3^{(b)} {\rm T}_2^{(c)}}}
\end{split}
\end{align}
explicitly. This exercise shows that reducible diagrams do contribute to the exponent in the multi-leg case, and on the face of it the non-Abelian exponentiation theorem does not generalise. The work of Ref.~\cite{Gardi:2010rn,Mitov:2010rp,Gardi:2011wa,Gardi:2011yz,Dukes:2013wa,Dukes:2013gea,Gardi:2013ita} shows that it does in fact generalise but in a rather non-trivial way. A few hints can already be drawn from the simple example above: first, it is convenient to consider together sets of diagrams which are related by permutation of the order of gluon attachments to the Wilson lines; we refer to the entire set as a single web. Second, the web as a whole yields a non-Abelian colour factor (in the present case we got 
${\rm i}f^{abc} {\rm T}_1^{(a)} {\rm T}_3^{(b)} {\rm T}_2^{(c)}$, which is the same as the colour factor of a three gluon vertex connecting to the three lines). Third, in the combination of integrals accompanying this colour factor in the exponent certain subdivergences conspire to cancel: while ${\cal F}_{(2a)}$ and ${\cal F}_{(2b)}$ each have a double pole, their difference, which enters the exponent via (\ref{two_loop_example_exp}), has a single pole.

All these properties can be formulated and proven in full generality~\cite{Gardi:2010rn,Mitov:2010rp,Gardi:2011wa,Gardi:2011yz,Dukes:2013wa,Dukes:2013gea,Gardi:2013ita}. 
Using a functional integral formalism~\cite{Laenen:2008gt,Gardi:2010rn} employing the replica trick of statistical physics~\cite{Replica} (see below) it was shown in ref.~\cite{Gardi:2010rn} that the exponent can be written as a sum of webs $W_i$, where each web is a set $\left\{D\right\}_i$ that comprises all the diagrams $D$ which are related to each other by permutations of the order of gluon emissions along each of the Wilson lines; a given diagram $D$ contributes with an ECF which is itself a linear combination of the ordinary colour factors of diagrams in the set. The resulting structure is then:
\begin{align}
S=\exp\left[\sum_i W_i\right]\,,
\qquad\qquad
W_i=\sum_{\left\{D\right\}_i}{\cal F}_{D}\, {\color[rgb]{1.0,0.0,0.0}\widetilde{C}_{D}}
=\,\sum_{\left\{D\right\}_i}{\cal F}_{D}\,
\sum_{\left\{D'\right\}_i}\,{\color[rgb]{0.0,0.0,1.0}R_{DD'}}\,\,C_{D'}\,= {\cal F}^T {\color[rgb]{0.0,0.0,1.0}R} C\,
\end{align}
where $R$ is a matrix whose entries are rational numbers. 
We refer to $R$ as the \emph{web mixing matrix}. The replica trick led to a general algorithm (and a combinatorial formula) to compute the matrix $R$ for a general web.
This matrix determines the way in which colour and kinematic information is entangled in the exponent. It has some remarkable properties: 
\begin{enumerate}
\item{} $R$ is idempotent, namely $R^2=R$, which implies that it acts as a projection operator on the space of colour factors or kinematic integrals of individual diagrams. The idempotence property was proven in~\cite{Gardi:2011wa} using the replica trick formalism.
Being idempotent, $R$ is diagonalisable, and its eigenvalues are just zeros and ones. Each left eigenvector of unit eigenvalue 
corresponds to a particular linear combination of colour factors that enters the exponent with a corresponding combination of kinematic integrals. Formulae for the number of such independent colour factors have been derived in refs.~\cite{Dukes:2013wa,Dukes:2013gea}. A basis for 
these colour factors was proposed in ref.~\cite{Gardi:2013ita}. 
%The multiplicity of the eigenvalue 1 (which is the rank $r$, or the trace, of the mixing matrix) 
%\begin{align}
%\begin{split}
%W\,=\, {\cal F}^{\mathrm T}\,\widetilde{C} 
%=&\,\,{\cal F}^{\mathrm T} \,R\, C\\
%=&\,\,\Big({\cal F}^{\mathrm T} Y^{-1}\Big) \, Y \,R\, Y^{-1}\, \Big( Y \,C\Big)\\
%=\,\,&\Big({\cal F}^{\mathrm T} Y^{-1}\Big) \, {\rm diag}(\lambda_1,\lambda_2,\ldots,\lambda_d)  \,\Big( Y \,C\Big)\\
%=\,\,&\sum_{H=1}^{r}\Big({\cal F}^{\mathrm T} Y^{-1}\Big)_H \, \,\Big( Y \,C\Big)_H
%\end{split}
%\end{align}
\item{} $R$ admits a zero-sum-row property: $\sum_{D'} R_{DD'}=0$ for all $D$. This corresponds to the fact that the fully symmetric colour factor is projected out. This property was proven in~\cite{Gardi:2011wa} using combinatorial methods.
\item{} $R$ admits a weighted zero-sum column property: $\sum_{D}s(D) R_{DD'}=0$, where $s(D)$ is a symmetry factor that counts the number of ways of shrinking subdiagrams to the origin in maximally reducible diagrams. This property has not yet been proven in general, but it relates to the cancellation of the leading subdivergences in the exponent. 
\end{enumerate}
The mixing matrix has a rich combinatorial structure which represents  detailed properties of soft-gluon exponentiation. Of special interest is the connection with the renormalizability of the Wilson-line vertex: in each web, diagrams are combined such that the leading subdivergences cancel, while all remaining multiple poles match the structure of commutators of lower-order webs~\cite{Mitov:2010rp,Gardi:2011yz}. The fact that webs renormalize independently will be crucial for computing them.

\subsection{The non-Abelian exponentiation theorem}

\begin{figure}[htb]
\begin{center}
\hspace*{-30pt}\begin{minipage}[b]{0.7\linewidth}
\scalebox{.6}{\includegraphics{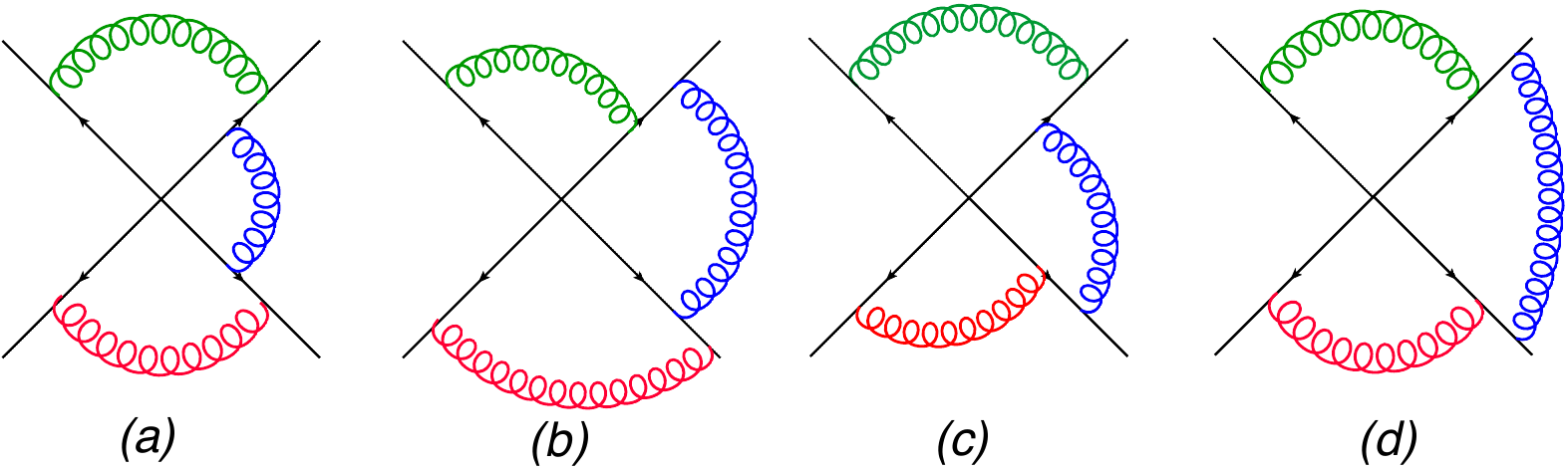}}
\end{minipage}
\begin{minipage}[b]{0.09\linewidth}
\begin{tabular}{c}
\scalebox{.3}{\includegraphics{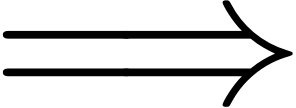}}
\\
\\
\\
\\
\\
\,
\end{tabular}
\end{minipage}
\begin{minipage}[b]{0.1\linewidth}
\begin{tabular}{c}
\scalebox{.48}{\includegraphics{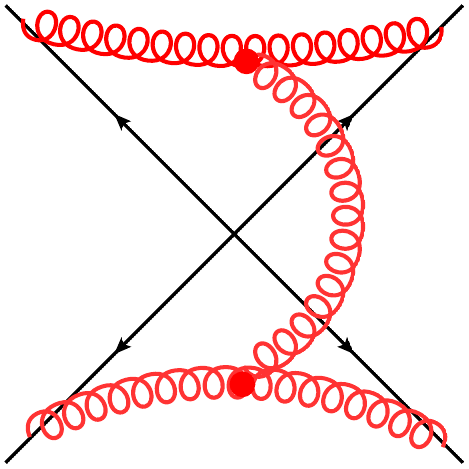}}
\\
\\
\\
\\
\\
\\
\end{tabular}
\end{minipage}

\vspace{-60pt}
\caption{Left: The four diagrams (a) through (d) (where four Wilson lines are connected by three gluons) forming the 1-2-2-1 web. Right: a graph representing the connected colour factor generated by this web. }
\label{fig:1221}
\end{center}
\end{figure}

The next step was taken in ref.~\cite{Gardi:2013ita} where it was shown in general that \emph{all the colour factors appearing in the exponent correspond to connected graphs}. This completes the generalization of the non-Abelian exponentiation theorem to the multi-line case. A three-loop example is shown in figure~\ref{fig:1221}.
The 1-2-2-1 web comprises four diagrams corresponding to all possible orderings of the gluon attachments along the lines. The contribution of this web to the exponent is
\begin{align}
\begin{split}
W_{1-2-2-1}&=\left(\begin{array}{c}{\cal F}_{(a)}\\{\cal F}_{(b)}\\{\cal F}_{(c)}\\{\cal F}_{(d)}\end{array}\right)^T\frac{1}{6}\left(\begin{array}{rrrr}1&-1&-1&1\\-2&2&2&-2\\-2&2&2&-2\\1&-1&-1&1\end{array}\right)\left(\begin{array}{c}C_{(a)}\\C_{(b)}\\C_{(c)}\\C_{(d)}\end{array}\right)
\\
&=\frac{1}{6}\Big({\cal F}_{(a)}-2{\cal F}_{(b)}-2{\cal F}_{(c)}+{\cal F}_{(d)}\Big){\color[rgb]{1.0,0.0,0.0}\Big(C_{(a)}-C_{(b)}-C_{(c)}+C_{(d)}\Big)}
\end{split}
\end{align}
where the combination of colour factors is $C_{(a)}-C_{(b)}-C_{(c)}+C_{(d)}=f^{abe}f^{cde}T_1^aT_2^bT_3^cT_4^d$, corresponding to the connected diagram on the r.h.s of figure~\ref{fig:1221}.

Having seen a couple of examples of how connected colour factors emerge (many more can be found in~\cite{Gardi:2013ita}) let us briefly summarise the main ideas behind the proof of the general theorem~\cite{Gardi:2013ita}. 
To this end let us first recall the replica trick formalism of~\cite{Gardi:2010rn}. The generating functional for all radiative corrections to $S$ in eq.~(\ref{S}), involving $L$ Wilson lines, is:
\begin{equation}
{\cal Z}=\int\left[ {\cal D}{A}^\mu\right]\,e^{{\mathrm i}S[A^\mu]}\,\left[{\Phi}^{(1)}\otimes {\Phi}^{(2)}\otimes\cdots \otimes {\Phi}^{(L)}\right]
\end{equation}
where a functional integral is taken over the gauge field $A^\mu$, the action is $S[A^\mu]$, and the Wilson lines are defined in eq.~(\ref{path-ordered_exp}) for $l=1..L.$ 
To obtain the Feynman rules corresponding to the exponent $w$ in eq.~(\ref{S}) we need to consider $\ln {\cal Z}$. This is achieved using the replica trick as follows: at the first step one replicates the theory $N$ times 
by introducing $N$ non-interacting replicas of the theory, each of which is sourced by each of the Wilson lines:
\begin{equation}
\label{replicated_Z}
{\cal Z}^{\color[rgb]{1.0,0.0,0.0}N}=\int \left[{\cal D}{A}_{{\color[rgb]{0.0,0.0,0.0}\mu}}^{\color[rgb]{1.0,0.0,0.0}1}\right]
\ldots\left[{\cal D}{A}_{{\color[rgb]{0.0,0.0,0.0}\mu}}^{\color[rgb]{1.0,0.0,0.0}N}\right] \,e^{{\mathrm i}\sum_{\color[rgb]{1.0,0.0,0.0}n} S[A_{{\color[rgb]{0.0,0.0,0.0}\mu}}^{\color[rgb]{1.0,0.0,0.0}{n}}]}\,\bigg[({\Phi}^{(1)}_{\color[rgb]{1.0,0.0,0.0}1}{\Phi}^{(1)}_{\color[rgb]{1.0,0.0,0.0}2}\ldots{\Phi}^{(1)}_{\color[rgb]{1.0,0.0,0.0}N})\otimes 
%({\Phi}^{(2)}_{\color[rgb]
%{1.0,0.0,0.0}1}{\Phi}^{(2)}_{\color[rgb]{1.0,0.0,0.0}2}\ldots{\Phi}^{(2)}_{\color[rgb]{1.0,0.0,0.0}N} )
%\otimes\cdots
\cdots \otimes( {\Phi}^{(L)}_{\color[rgb]{1.0,0.0,0.0}1}{\Phi}^{(L)}_{\color[rgb]{1.0,0.0,0.0}2}\ldots{\Phi}^{(L)}_{\color[rgb]{1.0,0.0,0.0}N})\bigg]
 \end{equation}
Diagrams in the replicated theory ${\cal Z}^N$ would have modified colour factors which depend on $N$ as we explain below. At the next step one expands in powers of $N$, 
\begin{equation}
{\cal Z}^N = 1+N\ln {\cal Z} +{\cal O}(N^2)\,.
\end{equation}
Expanding the replicated-theory results for a given diagram in powers of $N$ and identifying the coefficient of $N^1$ yields the contribution of the diagram to $\ln {\cal Z}$, namely to the exponent $w$.

Let us now explain how colour factors in the replicated theory become $N$-dependent. Consider one of the Wilson lines in the replicated generating functional (\ref{replicated_Z}), which by definition takes the form:
\begin{equation}
\label{phi_prod}
{\color[rgb]{0.0,0.0,0.0}\left[{\Phi}_{\color[rgb]{1.0,0.0,0.0}1}^{(l)}{\Phi}_{\color[rgb]{1.0,0.0,0.0}2}^{(l)}\ldots{\Phi}_{\color[rgb]{1.0,0.0,0.0}N}^{(l)}\right]_{a_1b_1}=
\left({\cal P}\exp\left[{\mathrm i}g_s\int dt\,{A}_{{\color[rgb]{1.0,0.0,0.0}1}}^{\color[rgb]{0.0,0.0,0.0}(l)}
(t)\right]\right)_{a_1c_2}
\ldots
\left({\cal P}\exp\left[{\mathrm i}g_s\int dt\,{A}_{{\color[rgb]{1.0,0.0,0.0}N}}^{\color[rgb]{0.0,0.0,0.0}(l)}(t)\right]\right)_{c_{N}b_1}}\,,
\end{equation}
where we explicitly displayed the replica and colour indices and used a short-hand notation for the gauge field of replica $n$ along the Wilson line $l$: ${A}_{{\color[rgb]{1.0,0.0,0.0}n}}^{\color[rgb]{0.0,0.0,0.0}(l)}(t) = \beta_l^{\mu}A_{{\color[rgb]{0.0,0.0,0.0}\mu}}^{\color[rgb]{1.0,0.0,0.0}{n}}(t\beta_l)$.
Note that the order of the colour generators (belonging to different replicas) in each diagram generated must be preserved; this can be formally represented by a replica-ordering operator ${\color[rgb]{0.0,0.0,1.0}{\cal R}}$ such that (\ref{phi_prod}) takes the form:
\begin{equation}
\label{rep_ordered}
{\color[rgb]{0.0,0.0,1.0}{\cal R}}{\cal P}\exp\left[{\mathrm i}g_s\sum_{n=1}^N\int dt{A}_{{\color[rgb]{0.0,0.0,0.0}n}}^{\color[rgb]{0.0,0.0,0.0}(l)}(t)\right]\,.
\end{equation}
Here, upon expanding the exponential, colour generators associated with the gauge field of a given replica are ordered according to the relative position of the field along the Wilson line (path ordered), while those belonging to different replicas are ordered according to their replica numbers (replica ordered).  At this point it is already clear that the colour factors in the replicated theory are modified through the operation of ${\color[rgb]{0.0,0.0,1.0}{\cal R}}$, making them $N$-dependent. This formulation was the basis of a Maple code~\cite{Gardi:2010rn} which allows to compute the web mixing matrices for any web. 

\begin{figure}[htb]
\begin{center}
\includegraphics[width=.7\textwidth]{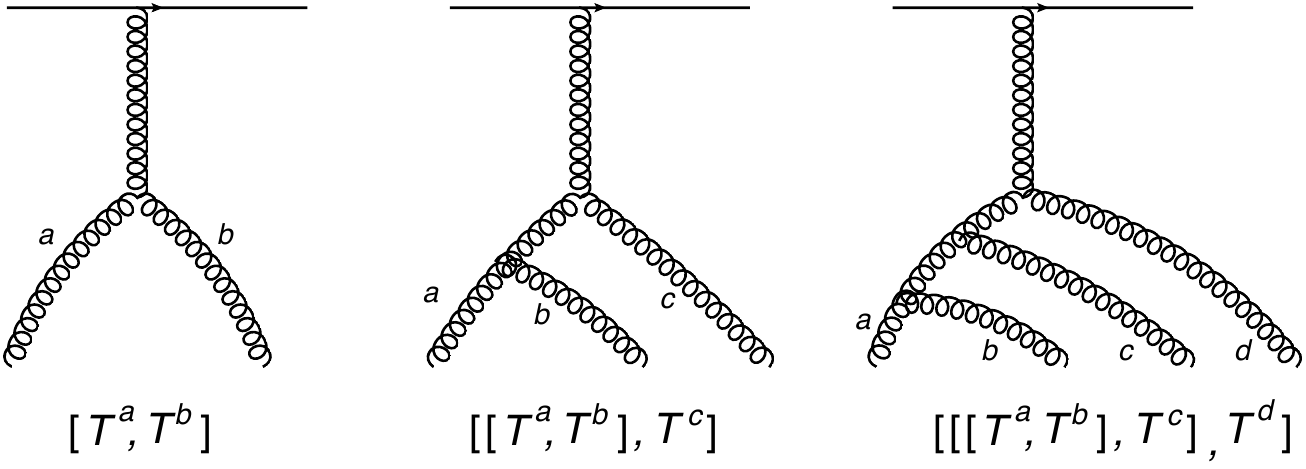}
\caption{Examples of colour factors arising in the exponent owing to the (repeated) application of the Baker-Campbell-Hausdorff formula to implement path ordering of fields belonging to the same replica and replica ordering otherwise. Connected graphs are obtained using the colour algebra. } \label{BCH_connected}
\end{center}
\end{figure}

\begin{figure}[htb]
\begin{center}
\includegraphics[width=.7\textwidth]{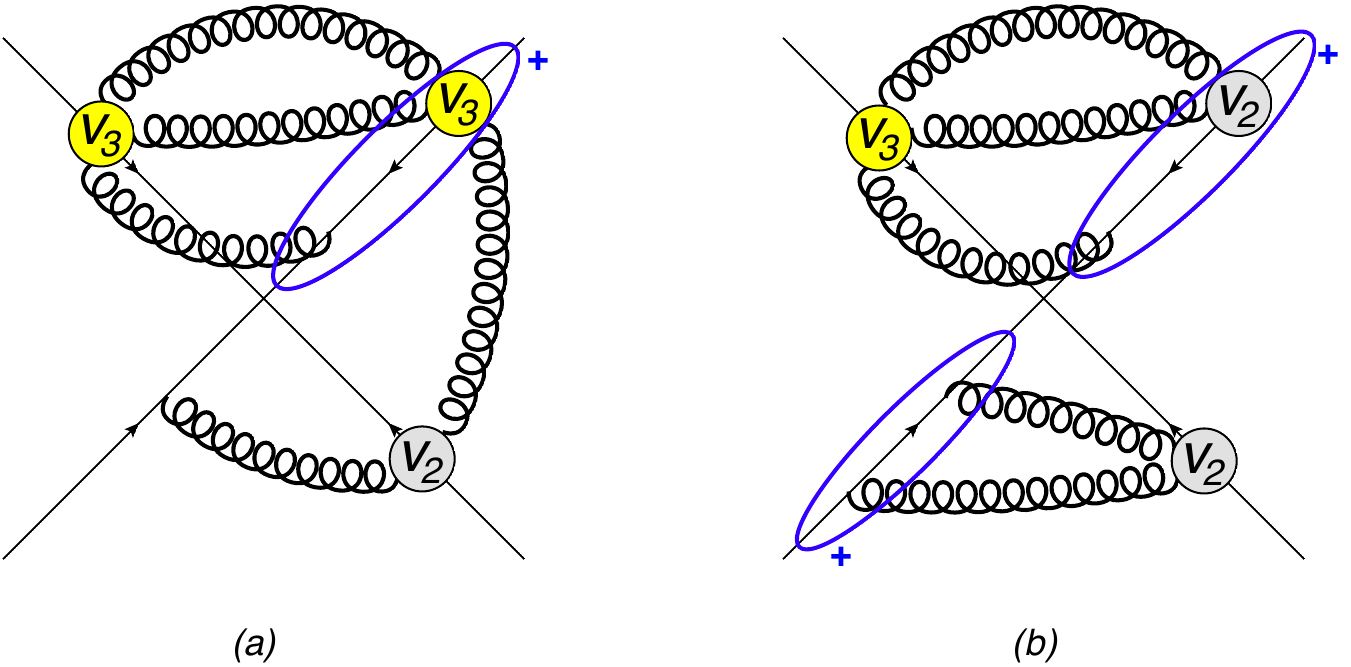}
\caption{Examples of graphs made out of the effective vertices $V_n$. The $+$ sign encircling vertices on the same Wilson line is there to indicate that a symmetric combination must be taken (no path-ordering is required).  } \label{Vn_connected}
\end{center}
\end{figure}

Equivalently, eq.~(\ref{rep_ordered}) can be converted into an ordinary exponential upon repeated use of the Baker-Campbell-Hausdorff formula to implement the path- and replica-orderings: 
\begin{align}
\label{BCH}
\begin{split}
\exp\left[({\mathrm i}g_s) \int_0^{\infty} ds \sum_{i}A_i(s)
 +\frac12 ({\mathrm i}g_s)^2 \int_0^{\infty} ds dt 
\Bigg\{\theta(s>t) \sum_{i}\left[A_i(s),A_i(t)\right]+\sum_{i<j}\left[A_i(t),A_j(s)\right]\Bigg\}
+\ldots\right]\,.
\end{split}
\end{align}
In this formulation correlated emission of $n$ gluons from the Wilson line may be realised though an effective vertex $V_n$ involving a nested commutator of $n$ gauge fields. This vertex is non-local in configuration space (e.g. the commutator of two fields in (\ref{BCH}) involves the two positions $t$ and $s$) but importantly its colour factor, originating from a fully-nested commutator, corresponds to a fully connected $n$-gluon emission graph such as those shown in figure~\ref{BCH_connected}.
Having established the effective Feynman rules, the proof in~\cite{Gardi:2013ita} proceeded by using the usual replica trick argument: out of all the graphs that can be formed using the effective vertices $V_n$ on the $L$ Wilson lines, only those which are fully connected (such as the one on figure~\ref{Vn_connected} (a)) have a linear dependence on the number of replicas $N$. 
In particular graphs with two separate subgraphs, as in figure~\ref{Vn_connected} (b), scale at least as $N^2$, since the replicas can be assigned independently in the two subgraphs.
Consequently only connected graphs made of the vertices $V_n$ contribute to the $\ln {\cal Z}$ theory. Given that the colour factor of each vertex is itself connected, all graphs in the $\ln {\cal Z}$ theory -- namely in the exponent -- have connected colour factors.  
This completes our review of the non-Abelian exponentiation theorem, and we are ready to consider the calculation of corresponding integrals in the next section. 

%%%%%%%%%%%%%%%%%%%%%%%%
\section{From webs to polylogs}

\begin{figure}[htb]
\begin{center}
\scalebox{.7}{\includegraphics{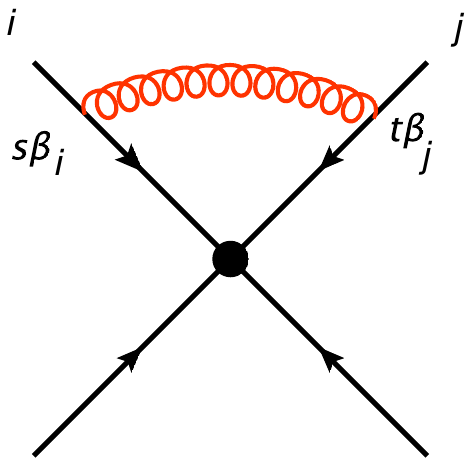}}
\caption{One loop web, where the gluon is emitted between Wilson lines $i$ and $j$
%, whose kinematic part is given by eq.~(\ref{eq:Fijoneloop2})
.
}
\label{1loopfig}
\end{center}
\end{figure}

\begin{figure}[htb]
\begin{center}
\scalebox{.9}{\includegraphics{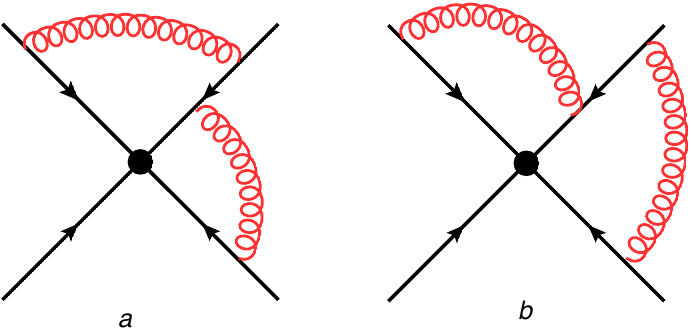}}
\hspace*{50pt}
\scalebox{0.6}{\includegraphics{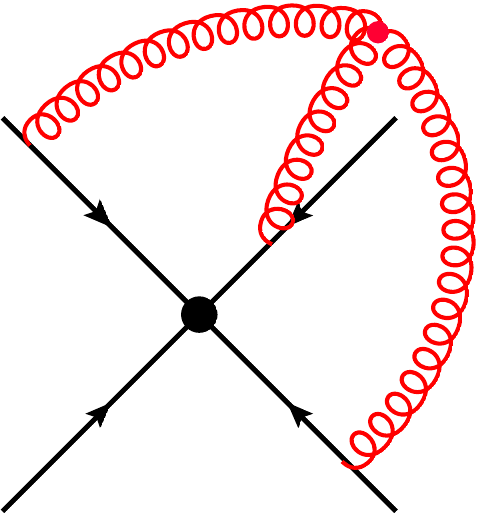}}
\caption{Two-loop graphs connecting three Wilson lines. Left: The two diagrams, called respectively (a) and (b), form together a 1-2-1 web, which we denote by $w_{121}^{(2)}$. Right: A connected web correlating the three Wilson lines. These two webs have the same colour factor.}
\label{2loopfig}
\end{center}
\end{figure}

\begin{figure}[htb]
\begin{center}
\scalebox{.9}{\includegraphics{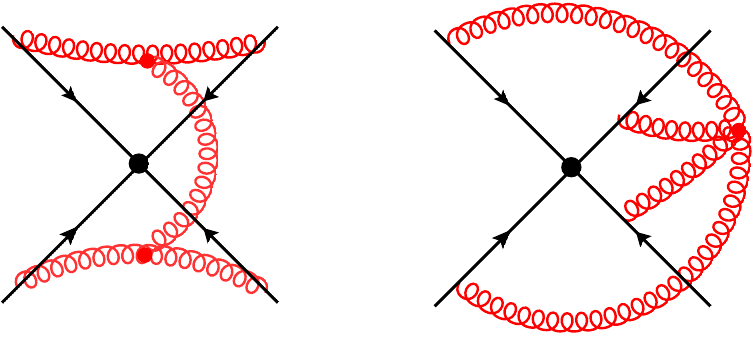}}
\caption{Connected three-loop diagrams with 3- or 4-gluon vertices spanning four Wilson lines.}
\label{Connected_four}
\end{center}
\end{figure}

\begin{figure}[htb]
\begin{center}
\scalebox{.9}{\includegraphics{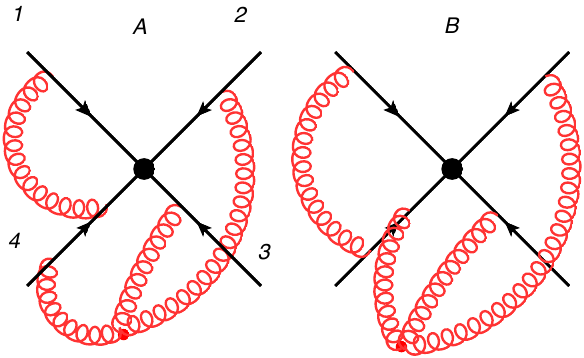}}
\caption{The 1-1-1-2 web. Each of the two diagrams in this web ($A$ and $B$) contains two connected subdiagrams, one of which has a three-gluon vertex.}
\label{two_pieces}
\end{center}
\end{figure}

\begin{figure}[htb]
\begin{center}
\scalebox{0.7}{\includegraphics{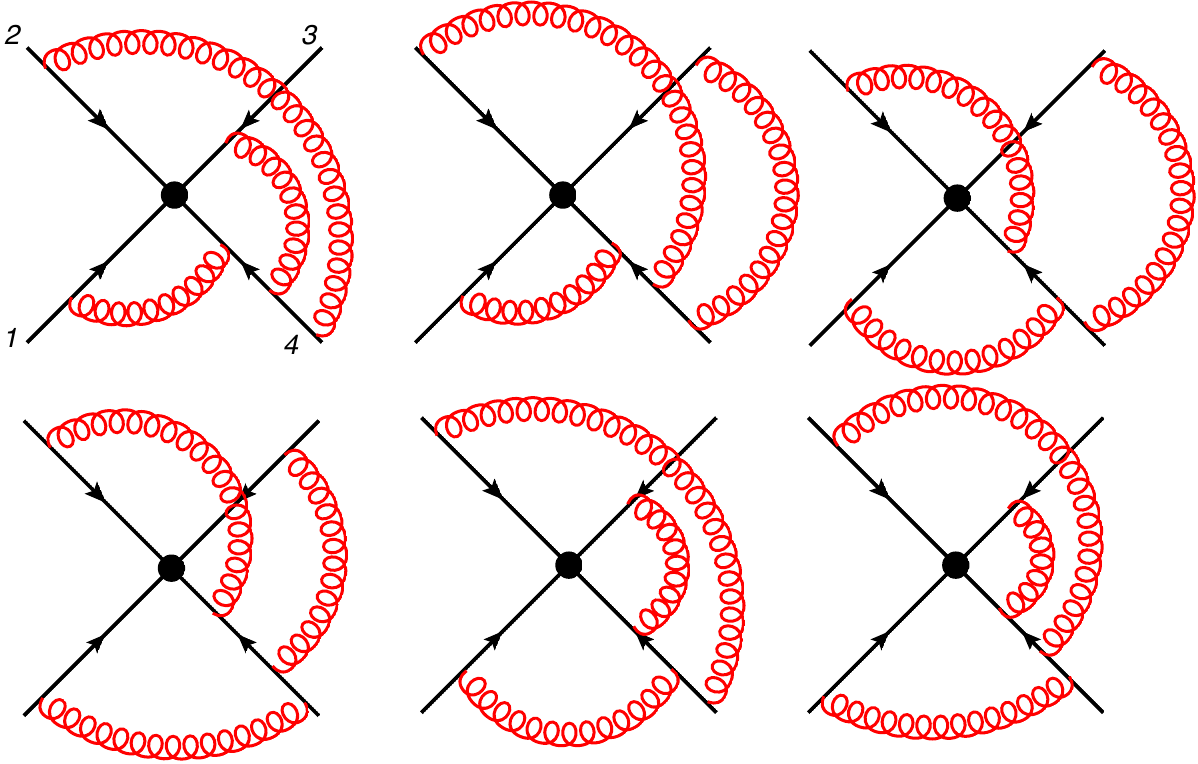}}
\caption{The six 3-loop diagrams forming the 1-1-1-3 web in which four eikonal lines are linked by three gluon exchanges. }
\label{1113}
\end{center}
\end{figure}

The soft anomalous dimension 
\begin{equation}
\label{Gamma_def}
\Gamma=-Z^{-1}\frac{dZ}{d\ln \mu}=\Gamma^{(1)}\alpha_s+\Gamma^{(2)}\alpha_s^2+\Gamma^{(3)}\alpha_s^3+\ldots
\end{equation}
 encodes the long-distance singularities of any scattering amplitude. The one-loop correction $\Gamma^{(1)}$ correlates pairs of Wilson lines only (figure~\ref{1loopfig}), and is therefore determined in terms of the angle-dependent cusp anomalous dimension\footnote{The cusp anomalous dimension itself has been known to two loops for many years~\cite{Korchemsky:1987wg}.}. 
The complete two-loop result $\Gamma^{(2)}$ correlating three Wilson lines (the relevant webs are shown in figure~\ref{2loopfig}) has been known for a few years now for both the massless and the massive cases~\cite{Aybat:2006wq,Aybat:2006mz,Kidonakis:2009ev,Mitov:2009sv,Becher:2009kw,Beneke:2009rj,Czakon:2009zw,Ferroglia:2009ep,Ferroglia:2009ii,Chiu:2009mg,Mitov:2010xw,Ferroglia:2010mi,Bierenbaum:2011gg}.
In the massless case the three-line correlation vanishes and the result for the soft anomalous dimension reduces into a sum over colour dipoles. Indeed, it has been shown that in the massless case factorization and rescaling symmetry lead to all-order constraints, suggesting a simple ansatz for the anomalous dimension in the form of a sum over dipoles~\cite{Becher:2009cu,Gardi:2009qi,Becher:2009qa,Dixon:2009gx,Dixon:2009ur}.
The first possible correction to this dipole formula may appear at three-loop order from diagrams involving four Wilson lines. Despite recent progress~\cite{Dixon:2008gr,Gardi:2009qi,Dixon:2009gx,Becher:2009cu,Becher:2009qa,Dixon:2009ur,Gardi:2009zv,Gehrmann:2010ue,Bret:2011xm,DelDuca:2011ae,Ahrens:2012qz,Naculich:2013xa,Caron-Huot:2013fea},  
it seems that general considerations fall short of excluding or fixing these corrections\footnote{A very interesting argument has been formulated recently~\cite{Caron-Huot:2013fea} based on the Regge limit, indicating that the dipole formula should be violated at four loops.}; further input from explicit calculations is needed. This provides extra motivation to compute $\Gamma$ at three loops, and specifically to determine its component involving four Wilson lines $\Gamma^{(3)}_4$. The relevant webs are those having colour factors of the form $f^{abe}f^{cde}T_1^aT_2^bT_3^cT_4^d$ and permutations. There are several webs that contribute: the most obvious ones are the connected three-loop webs of figure~\ref{Connected_four}, but there are others, which consist of two connected pieces (figure \ref{two_pieces}) and ones that involve three individual gluon exchanges: the 1-2-2-1 web of figure~\ref{fig:1221} and the 1-1-1-3 web of figure~\ref{1113}. 

While complete four-leg three-loop computations of scattering amplitudes are beyond the state of the art, determining the singularities may be possible. The techniques needed in order to evaluate the integrals for connected webs~\cite{Almelid:2013tb} as in figure~\ref{Connected_four} are rather different from those needed to evaluate the contributions of gluon exchange diagrams such as those of  figures~\ref{fig:1221} and~\ref{1113}. 
In the latter case a difficulty arises due to the fact that the overall degree of divergence is high, with some of the diagrams having ${\cal O}(1/\epsilon^3)$ singularities\footnote{Note that in a direct lightlike calculation there are additional collinear singularities and the leading pole is ${\cal O}(1/\epsilon^6)$.}, while determining the anomalous dimension requires to compute the ${\cal O}(1/\epsilon)$ pole, as we shall see below. Here the direct computation of the exponent in terms of webs, as described in section~\ref{non-abelian_exp_theorem}, becomes crucial. This will be demonstrated in what follows, where we review the recent progress in evaluating multiple-gluon-exchange webs~\cite{Gardi:2013saa}. 
%\begin{figure}[htb]
%\begin{center}
%\scalebox{1.0}{\includegraphics{4lines_many_loops.pdf}}
%\caption{An example multiple-gluon-exchange diagram connecting four semi-infinite Wilson lines at seven loops. The 4-velocities associated with the directions of the Wilson lines in Minkowki space are indicated by~$\beta_i$. The lines all meet at the origin, where there is a local effective vertex representing the hard interaction.  }
%\label{multi}
%\end{center}
%\end{figure}

\subsection{Multiplicative renormalizability and the soft anomalous dimension}

Before discussing the web integrals themselves, let us recall how the soft anomalous dimension is computed in terms of webs. $\Gamma$ is defined in (\ref{Gamma_def}) as the logarithmic derivative of the renormalization factor $Z$ of the product of Wilson lines $S=\exp[w]$, where $Z$ is a multiplicative renormalization factor defined so as to absorb the ultraviolet singularities of $S$ such that 
\begin{equation}
\label{mul_ren}
\exp[w] \, Z={\cal O}(\epsilon^0)\,. 
\end{equation}
To determine the ultraviolet singularities of $w$ in dimensional regularization we must introduce another infrared regulator; we note that $\Gamma$ itself is independent of this regulator, while intermediate results may depend on it.
With a regulator in place we get well-defined results for webs diagrams contributing to $w$ in $4-2\epsilon$ space-time dimensions, where $\epsilon>0$. Using (\ref{mul_ren}) along with the definition of $\Gamma$ we may express the perturbative coefficients of $\Gamma$ in terms of those of $w=\sum w^{(n,k)}\alpha_s^n\epsilon^k$:
\begin{subequations}
\label{gamma_w}
\begin{align}
\Gamma^{(1)}&=-2w^{(1,-1)}\,,\\
\Gamma^{(2)}&=-4w^{(2,-1)}-2\left[w^{(1,-1)},w^{(1,0)}\right]\,,\label{Gamma_2}\\
\begin{split}
\label{Gamma_3}
\Gamma^{(3)}&=-6w^{(3,-1)}+\frac32b_0\left[w^{(1,-1)},w^{(1,1)}\right]
+3\left[w^{(1,0)},w^{(2,-1)}\right]
+3\left[w^{(2,0)},w^{(1,-1)}\right]
\\&
+\left[w^{(1,0)},\left[w^{(1,-1)},w^{(1,0)}\right]\right]
-\left[w^{(1,-1)},\left[w^{(1,-1)},w^{(1,1)}\right]\right]\,,
\end{split}
\end{align}
\end{subequations}
where $b_i$ are the coefficients of the $\beta$ function. Commutators appear at two-loops and beyond as a consequence of the non-commutativity of the webs in $\exp[w]$ and the counterterms in $Z$. It is important to note that in contrast to the two-line case multiple 
poles do arise in the exponent~$w$. The structure dictated by multiplicative renormalizability implies that all such multiple poles correspond to commutators of lower orders, and can thus be predicted at order $n$ without computing ${\cal O}(\alpha_s^n)$ webs. This in turn provides a strong check of multi-loop calculations if these are organised as webs: at each order all but the single $1/\epsilon$ coefficient (which contribute to the anomalous dimension (\ref{gamma_w})) are predetermined.
This can be most concisely summarised by expressing the exponent of $Z$ in terms of the coefficients of the anomalous dimension $\Gamma^{(n)}$, namely
\begin{align}
\label{Z_exp}
\begin{split}
Z=&\,\exp\left\{
\,\frac{1}{2\epsilon}\,\Gamma^{(1)}\,\alpha_s
+\,\left(\frac{1}{4\epsilon}\,\Gamma^{(2)}-\frac{b_0}{4\epsilon^2}\,\Gamma^{(1)}\right)\,\alpha_s^2
\right.\\&
\left.
+\,\left(\frac{1}{6\epsilon}\,\Gamma^{(3)}
+\frac{1}{48\epsilon^2}\left[\Gamma^{(1)},\Gamma^{(2)}\right]-\frac{1}{6\epsilon^2}
\left(b_0\Gamma^{(2)}+b_1\Gamma^{(1)}\right)+\frac{b_0^2}{6\epsilon^3}\Gamma^{(1)}\right)\,\alpha_s^3
\,+\,{\cal O}(\alpha_s^4)\,
 \right\}\,,
\end{split}
\end{align}
where we note in particular that, at the three-loop order, ${\cal O}(\epsilon^{-2})$ singularities do occur in the exponent (even in a conformal theory where $b_n=0$) and these correspond to a commutator of $\Gamma^{(1)}$ with $\Gamma^{(2)}$. In explicit three-loop calculations the double poles can be identified in webs that have subdivergences: this was illustrated for the 1-2-3 and 1-1-1-3 webs in Ref.~\cite{Gardi:2011yz}. 

\subsection{One-loop calculation}

We are now ready to discuss the computation of the multi-gluon-exchange web integrals following~\cite{Gardi:2013saa}, and we start with the simplest example: the one-loop graph of figure~\ref{1loopfig} with a single gluon exchange between Wilson lines $i$ and $j$ with velocities $\beta_i$ and $\beta_j$. 
This is the familiar cusp configuration, and owing to rescaling invariance the result can only depend on $\gamma_{ij}=2\beta_i\cdot \beta_j/\sqrt{\beta_i^2\beta_j^2}$.
We use configuration space Feynman rules where the integration is over two parameters $s$ and $t$ corresponding to the positions of the vertices along the two Wilson lines. The expression for the web diagram is then:
\begin{align}
\label{w1}
\begin{split}
w^{(1)}
  &=T_i\cdot T_j \,\mu^{2\epsilon} g_s^2\,{\cal N}\,\beta_i\cdot\beta_j\!
\int_0^{``\infty"}\!\! ds\! \int_0^{``\infty"} \!\! dt
  \, \Big(-(s\beta_i-t\beta_j)^2\Big)^{\epsilon-1}
%\\   &\quad \times 
%{\rm e}^{-m\left(s\sqrt{-\beta_i^2}+t\sqrt{-\beta_j^2}\right)}
\\
&=T_i\cdot T_j \,\kappa\,\gamma_{ij}\,
\int_0^{\infty} d\sigma \int_0^{\infty}  d\tau
  \Big(\sigma^2+\tau^2-\gamma_{ij} \sigma \tau\Big)^{\epsilon-1}
%\\   &\quad \times 
\,{\rm e}^{-\sigma-\tau}
\\
 &=
T_i\cdot T_j \, \kappa 
 \,\Gamma(2\epsilon) \,\int_0^1dx \, p_{\epsilon}(x,\alpha_{ij})
\end{split}
\end{align}
where ${\cal N}\equiv
\frac{\Gamma(1-\epsilon)}{4\pi^{2-\epsilon}}$ and $
\kappa\equiv -\left(\frac{\mu^2}{m^2}\right)^{\epsilon} \frac{g_s^2}{2}\,{\cal N}$ and
\begin{equation}
\label{pq}
p_{\epsilon}(x,\alpha_{ij})\,\equiv\,\gamma_{ij}\,\Big[q(x,\alpha_{ij})\Big]^{\epsilon-1}\,,\qquad 
q(x,\alpha_{ij})=x^2+(1-x)^2-x(1-x)\gamma_{ij}, \qquad  \gamma_{ij}=-\frac{1}{\alpha_{ij}}-\alpha_{ij}\,.
\end{equation}
In the first line of eq.~(\ref{w1}) we indicated that an infrared cutoff must be imposed using the notation $``\infty"$ for the upper integration limit. The cutoff is implemented in a systematic way through an exponential regulator which is made explicit in the second line. Exactly the same regulator will be applied to each of the gluons in a multi-gluon-exchange diagram. The result for the anomalous dimension will not depend on the regulator but, as we shall see, intermediate results at higher orders will.
In the second line of (\ref{w1}) we rescaled the parameters and arranged the kinematic dependence in terms of $\gamma_{ij}$. In the third line we changed the integration to be the total distance $\lambda=\sigma+\tau$ and $x=\sigma/(\sigma+\tau)$. The latter varies in the range $[0,1]$ and describes the gluon emission angle with respect to the Wilson lines. 
The integration over the total distance $\lambda$ yields the expected ultraviolet singularity. Finally, in performing the last integration, over the the gluon emission angle, it becomes apparent that the most convenient kinematic variable is $\alpha_{ij}$ which is related to $\gamma_{ij}$ as in eq.~(\ref{pq}).
\begin{figure}[htb]
\begin{center}
\scalebox{0.8}{\includegraphics{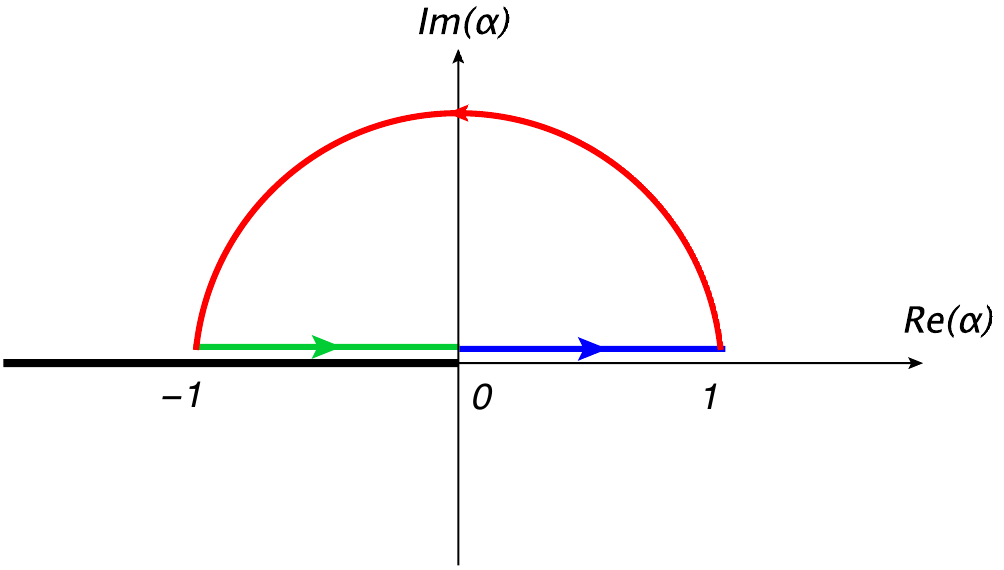}}
\caption{The analytic structure of the one-loop result in the complex $\alpha$ plane -- a logarithmic branch cut along the negative real axis -- shown together with a contour describing the values of $\alpha$ for real values of $\gamma$: the $\alpha\in (0,1)$ region corresponds to space-like kinematics (one incoming and one outgoing parton) where  $\gamma$ varies between $-\infty$ and $-2$;
next, the region of complex $\alpha$ with a positive imaginary part corresponds to the Euclidean region where $-2<\gamma<2$; and finally the region where $\alpha$ is near the branch cut, $\alpha=\alpha_r+{\rm i} \varepsilon$ with $\alpha_r\in (-1,0)$ and $\varepsilon>0$, corresponds to time-like kinematics.  }
\label{alpha_plane}
\end{center}
\end{figure}
This definition implies an inversion symmetry $\alpha_{ij}\to 1/\alpha_{ij}$, and we choose to consider values inside the unit circle, $|\alpha_{ij}|\leq 1$. At ${\cal O}(\alpha_s^1\,\epsilon^{-1})$ we get:
\begin{align}
\label{w1_1}
\begin{split}
w^{(1,-1)}(\alpha_{ij}) &=- \frac{T_i\cdot T_j }{4\pi}
  \int_0^1 {\rm d}x\ p_0(x,\alpha_{ij})\\
&= \frac{T_i\cdot T_j }{4\pi}\left(\alpha_{ij}+\frac{1}{\alpha_{ij}}\right)
  \int_0^1 {\rm d}x\frac{1}{x^2 + (1 - x)^2 + x(1 - x) (\alpha_{ij} + 1/\alpha_{ij})}
\\
&= -\frac{T_i\cdot T_j }{4\pi }\,\frac{1+\alpha_{ij}^2}{1-\alpha_{ij}^2}
  \int_0^1 {\rm d}x \left(\frac{1}{x - \frac{1}{1 - \alpha_{ij}}} - \frac{1}{x + \frac{\alpha_{ij}}{1 - \alpha_{ij}}} \right)
\\
&= -\frac{T_i\cdot T_j  }{2 \pi}\, r(\alpha_{ij})\,\ln\left({\alpha_{ij}}\right)
\end{split}
\end{align}
where the rational factor is $r(\alpha_{ij})\equiv\frac{1+\alpha_{ij}^2}{1-\alpha_{ij}^2}$. In a multi-gluon-exchange web such a rational factor will be associated with each gluon. The analytic structure of the result can be understood from physical considerations of the various kinematic limits. The physical regions are shown in figure~\ref{alpha_plane}.
Positive values of $\alpha_{ij}$ correspond to space-like kinematics as in deep inelastic scattering, with one parton incoming the other outgoing. The extremal value, $\alpha_{ij}=1$, corresponds to the forward limit or a straight infinite Wilson line. Here the zero of the logarithm cancels the pole of the rational factor.
Negative values of $\alpha_{ij}$ correspond to time-like kinematics where both partons are in the final state. Here the result should have an imaginary part, and it does. The analytic continuation of $\alpha_{ij}$ is through the upper half plane, such that $\alpha_{ij}\to \alpha_{ij}+{\rm i}0$. The extremal value here is $\alpha_{ij}=-1$; this corresponds to heavy quark production near threshold, where the velocity of the produced quarks is very small. Note that the rational factor $r(\alpha_{ij})$ generates the expected Coulomb singularity. Finally, the $\alpha_{ij}\to 0$ corresponds to the lightlike limit; here the rational factor tends to one, and there is a logarithmic divergence owing to the collinear singularity. 
Clearly the physical region for $\alpha_{ij}$ in figure~\ref{alpha_plane} is exactly the same at higher orders. Specifically, the only special points where one would expect the function to have branch points are $\alpha_{ij}=\left\{0,\pm 1,\infty\right\}$. In a multi-gluon-exchange web there will be dependence on several $\alpha_{ij}$ variables, but for each of these one still expects the same physical region and the same set of branch points.

\subsection{Two-loop calculation and the subtracted web}

Moving on to two loops, we can gather further understanding of the analytic structure of gluon-exchange webs. Calculations of the two-loop graphs of figure~\ref{2loopfig} have already been done in \cite{Ferroglia:2009ii,Mitov:2010xw}.  We will not reproduce any of the calculations here, but rather review some interesting properties of the result identified in Ref.~\cite{Gardi:2013saa}. The 1-2-1 web (the two diagrams on the left hand side in figure~\ref{2loopfig})  is given by
\begin{align}
\begin{split}
w_{121}^{(2,-1)}(\alpha_{ij},\alpha_{jk})
=&
 -{\rm i}f^{abc} T_i^aT_j^bT_k^c
\, \frac{1}{(4\pi)^2}\,
r(\alpha_{ij})r(\alpha_{jk})
\Big(\ln(\alpha_{ij})S_1(\alpha_{jk})- \ln(\alpha_{jk})S_1(\alpha_{ij})\Big)\,,
\end{split}
\end{align}
where 
\begin{align}
S_1(\alpha)=-2\,\frac{1}{r(\alpha)}\, \int_0^1 dx \,p_0(x,\alpha)\, \ln(x) =-2\left[-2 {\rm Li}_2(\alpha)+\frac12 \ln^2(\alpha)-2 \ln(1-\alpha) \ln(\alpha)+2\zeta_2\right]\,.
\end{align}
Our expectations with regards to the position of the branch points are clearly fulfilled. A convenient way to summarise the analytic structure of polylogarithmic functions is to consider their symbol~\cite{Goncharov:arXiv0908.2238,Goncharov:2010jf,Duhr:2011zq,Duhr:2012fh}. The symbol of the function $S_1(\alpha)$ is: 
\begin{align}
{\cal S}\left[S_1(\alpha)\right] &=4 \ln \alpha\otimes\ln  (1-\alpha)-2 \ln \alpha\otimes \ln \alpha\,,
\end{align}
where we use the convention used for the co-product term $\Delta_{1,1,\ldots,1}$ \cite{Duhr:2012fh} writing the $\ln$ explicitly.

According to eq.~(\ref{Gamma_2}) the anomalous dimension receives contributions from three terms (see eq.~(\ref{Gamma_2_explicit})): the two webs of figure~\ref{2loopfig}, as well as a commutator of single-gluon integrals, the latter taking the form:
\begin{align}
\begin{split}
\left[w^{(1,-1)},w^{(1,0)}\right]=-{\rm i} f^{abc} T_i^a T_j^b T_k^c\, \frac{1}{(4\pi)^2}\,
  r(\alpha_{ij}) \, r(\alpha_{jk})
\,4\Big(\ln(\alpha_{ij}) R_1(\alpha_{jk})-\ln(\alpha_{jk}) R_1(\alpha_{ij})\Big)\,,
\end{split}
\end{align}
where
\begin{align}
R_1(\alpha)=\frac{1}{2}\,\frac{1}{r(\alpha)}\,\int_0^1dx \, p_0(x,\alpha)\, \ln q(x,\alpha)=2{\rm Li}_2(-\alpha)+2\ln(\alpha)\ln(1+\alpha)-\frac12\ln^2(\alpha)+\zeta_2\,,
\end{align}
and its symbol is:
\begin{align}
{\cal S}\left[R_1(\alpha)\right] &=2 \ln \alpha\otimes \ln (1+\alpha)-\ln \alpha\otimes \ln \alpha\,.
\end{align}
Given that the commutator has the same rational prefactor as the 1-2-1 web it is natural to combine the two, yielding the following expression for the anomalous dimension:
\begin{align}
\label{Gamma_2_explicit}
\begin{split}
\Gamma^{(2)}&=-4w^{(2,-1)}_{3g}\underbrace{-4w_{121}^{(2,-1)}-2\left[w^{(1,-1)},w^{(1,0)}\right]}_{\displaystyle -4{\overline{w}_{121}^{(2,-1)}}}
\end{split}
\end{align}
where the combination $\overline{w}_{121}^{(2,-1)}$ is referred to as the 1-2-1 \emph{subtracted web}. More generally a subtracted web is defined as the web plus the corresponding contribution of commutators of its subdiagrams to the anomalous dimension. 
The 1-2-1 subtracted web is given by
\begin{align}
\label{w_sub}
\begin{split}
{\color[rgb]{0.0,0.0,0.0}\overline{w}_{121}^{(2,-1)}}&=- {\rm i}f^{abc} T_i^aT_j^bT_k^c 
\, \frac{1}{(4\pi)^2}\,
\,r(\alpha_{ij})\,r(\alpha_{jk})\Big(\ln(\alpha_{ij})U_1(\alpha_{jk})- \ln(\alpha_{jk})U_1(\alpha_{ij})\Big)
\end{split}
\end{align}
where the transcendental functions combine as 
\begin{equation}
U_1(\alpha)=S_1(\alpha)+2R_1(\alpha)=-2{\rm Li}_2(1-\alpha^2)-2 \ln^2(\alpha)\,.
\end{equation}
The symbol of $U_1$ is
\begin{align}
{\cal S}\left[U_1(\alpha)\right]&=-4\ln \alpha\otimes \ln \frac{\alpha}{1-\alpha^2}\,.
\end{align}
Here comes a very important observation\footnote{It should be pointed out that a related observation was recently made in refs.~\cite{Henn:2012qz,Henn:2012ia,Henn:2013wfa}. It was found there that the angle-dependent cusp anomalous dimension in ${\cal N}=4$ supersymmertic Yang-Mills can be expressed as a function of $\alpha^2$ through three loops, and at least for multiple-gluon-exchange diagrams this persists through six loops.  }: at symbol level we have a symmetry under $\alpha\to-\alpha$.
It should be emphasised that the separate contributions of the \hbox{1-2-1} web and the commutator do not have such a symmetry; only the combined expression does.  What \hbox{$\alpha\to -\alpha$} means physically is crossing a particle from the initial to the final state (recall the definition of $\gamma_{ij}$ and eq.~(\ref{pq})). So this is a relation between space-like and time-like kinematics. To understand it better, let us look at $\alpha_{ij}$ expressed in terms of dimensionful kinematic variables\footnote{I would like to thank Lance Dixon for illuminating discussions on this subject.}: 
\begin{align}
 \alpha_{ij}=\frac{\sqrt{1-\frac{\sqrt{m_i^2m_j^2}}{p_i\cdot p_j}}-\sqrt{1+\frac{\sqrt{m_i^2m_j^2}}{p_i\cdot p_j}}}{\sqrt{1-\frac{\sqrt{m_i^2m_j^2}}{p_i\cdot p_j}}+\sqrt{1+\frac{\sqrt{m_i^2m_j^2}}{p_i\cdot p_j}}}\,.
\end{align}
This relation becomes simple upon expansion near the lightlike limit. 
For ${m_i^2\to0}$ we get:
\begin{align}
\alpha_{ij}=\frac{\sqrt{m_i^2m_j^2}}{-2p_i\cdot p_j}\,\left[1+{\cal O}\left(\frac{m_i^2m_j^2}{(2p_i\cdot p_j)^2}\right)\right]
\end{align}
where the corrections appear as integer powers of $\frac{m_i^2m_j^2}{(2p_i\cdot p_j)^2}$.
Thus only the leading square-root term is sensitive to the relative sign of the momenta $p_i$ and $p_j$, namely to whether a given particle is incoming or outgoing. We expect logarithmic branch cuts starting at $\alpha_{ij}=\left\{0,\pm 1,\infty\right\}$, namely entries 
of $\ln \alpha$, $\ln(1-\alpha)$ and $\ln(1+\alpha)$ in the symbol. 
Let us now examine the expansion of these entries near the lightlike limit. Entries of the form $\ln \alpha$ are expected: they present a logarithmic collinear singularity at small $\alpha$ and generate ${\rm i}\pi$ terms for timelike kinematics (both $p_i$ and $p_j$ are outgoing); at the symbol level, ${\cal S}(\ln (-\alpha))={\cal S}(\ln \alpha)$, so there the symmetry $\alpha\to-\alpha$ clearly holds. Entries of the form $\ln (1\pm \alpha_{ij})$, on the other hand, are proportional to the \emph{square-root of the squared masses} at leading order, violating the known analytic properties of amplitudes. Such square roots will be avoided if, and only if, all $\ln (1-\alpha)$ and $\ln (1+\alpha)$ entries conspire to combine into the combination $\ln (1-\alpha^2)$. This is exactly what happened in the subtracted-web combination $\overline{w}_{121}^{(2,-1)}$. 
Based on the expected analytic dependence on $m_i^2$, we expect this to apply for the contribution of any web to the anomalous dimension. Ref.~\cite{Gardi:2013saa} showed that the $\alpha\to -\alpha$ symmetry is broken for non-subtracted webs (and for individual commutators entering the anomalous dimension) due to the infrared regulator, but it is recovered for the subtracted web (at symbol level).

\subsection{General structure of multi-gluon-exchange webs and first three-loop results} 

Having deduced that organising the calculation in terms of subtracted webs is useful and leads to recovering the crossing symmetry $\alpha\to-\alpha$ at the symbol level, and furthermore that the symbol alphabet for subtracted multi-gluon-exchange webs is conjectured to be $\left\{\ln \alpha,\, \ln(1-\alpha^2)\right\}$, Ref.~\cite{Gardi:2013saa} went on to analyse the general form of the integrals for multi-gluon-exchange webs. 

Using the change of variables we used in the one-loop calculation for any gluon in an $n$ gluon exchange diagram we arrive at the following general form:
\begin{align}
\label{mge_kin}
\begin{split}
 {\cal F}^{(n)}&\sim\kappa^n
\Gamma(2n\epsilon)\int dx_1 dx_2\ldots dx_n\, \phi_{n-1}(x_1,x_2,\ldots, x_n;\epsilon)\, \prod_{k=1}^n p_{\epsilon}(x_k,\alpha_k) 
\\
&= \kappa^n \Gamma(2n\epsilon) \left(\prod_{k=1}^{n} \, r(\alpha_k) \right) \, s_n(\{\alpha_k\};\epsilon)
\end{split}
\end{align}
where $p_{\epsilon}(x_k,\alpha_k)$ is defined in (\ref{pq}) and $\phi_{n-1}$ is a transcendental function of uniform weight $n-1$, which is obtained along with the overall singularity $\Gamma(2n\epsilon)$ upon integrating over the distance variables of the $n$ gluons.
Performing the integrals over the gluon emission angles $x_k$, as in the second line of (\ref{mge_kin}), yields the familiar rational factor $r(\alpha_{ij})$ for each gluon exchange times a rather complicated transcendental function $s_n$ which can be expressed in terms of Goncharov multiple-polylogarithm. Fortunately, there is no need to perform this integral for individual diagrams, and the result is much more transparent when organised in the form of subtracted webs. Ref.~\cite{Gardi:2013saa} has shown that the integrals can be combined and the general result for a given subtracted web reads:
\begin{align}
\label{subtracted_web_mge_kin}
\begin{split}
 \overline{w}^{(n,-1)}= \left(\frac{\alpha_s}{4\pi}\right)^n \, C_{i_1,i_2,\ldots i_{n+1}}\,&
 \int dx_1 dx_2\ldots dx_n\, \times\,\prod_{k=1}^n p_0(x_k,\alpha_k) \,\times \,\\&{\cal G}_{n-1}\Big(x_1,x_2,\ldots, x_n;
q(x_1,\alpha_1), q(x_2,\alpha_2), \ldots q(x_n,\alpha_n)
\Big)\, ,
\end{split}
\end{align}
where $C_{i_1,i_2,\ldots i_{n+1}}$ is a connected colour factor involving the generators of up to $(n+1)$ Wilson lines and ${\cal G}_{n-1}$ is a polylogarithmic function of uniform transcendental weight $(n-1)$.
The contributions to the anomalous dimension need to be summed over all subtracted webs of order $n$, $\Gamma^{(n)}=-2n \,\sum_i \overline{w}^{(n,-1)}_i$ (where we discarded running coupling terms).

It was further argued, based on the alphabet conjecture and the form of eq.~(\ref{mge_kin}) that ${\cal G}_{n-1}$ is a sum of products of \emph{logarithms} of its arguments (the basic reasoning is that if ${\cal G}_{n-1}$ has polylogarithms of ratios of its argument the result would necessarily involve a richer alphabet). As a consequence $\overline{w}^{(n,-1)}$ takes the form of a sum of products of polylogarithms each involving a single cusp angle. Moreover, the basis of integrals can be systematically constructed, and this was so far done  for the class of four-line three-loop webs, where there are only two new functions $U_2(\alpha)$ and $\Sigma_2(\alpha)$.
Rather than quoting the result for the functions themselves (these can be found in~\cite{Gardi:2013saa}) we just quote their symbols, which are remarkably simple: 
\begin{align}
{\cal S}\left[U_2(\alpha)\right]&=4\ln \alpha\otimes \ln \frac{\alpha}{1-\alpha^2} \otimes \ln \frac{\alpha}{1-\alpha^2}\,,\\
{\cal S}\left[\Sigma_2(\alpha)\right]&=2\ln \alpha\otimes \ln \alpha\otimes \ln \alpha\,.
\end{align}
In terms of these functions the result for the 1-2-2-1 subtracted web of figure ~\ref{fig:1221} is
\begin{align}
\begin{split}
\overline{w}^{(3,-1)}_{(122334)}&=-\frac16 f^{abe}f^{cde}T_1^aT_2^bT_3^cT_4^d\,\,
\left(\frac{1}{4\pi}\right)^3\,
G(\alpha_{12},\alpha_{23}, \alpha_{34})\,
\end{split}
\end{align}
with
\begin{align}
\label{G1221_sym}
\begin{split}
&G(\alpha_{12},\alpha_{23}, \alpha_{34})=r(\alpha_{12})\,r(\alpha_{23})\,r(\alpha_{34})\, \Bigg[-8 U_2(\alpha_{12})\, \ln \alpha_{23}\, \ln \alpha_{34}
-8 U_2(\alpha_{34}) \,\ln \alpha_{12}\, \ln \alpha_{23}
\\&\hspace*{30pt}+16\Big(U_2(\alpha_{23})-2\Sigma_2(\alpha_{23})\Big) \,\ln \alpha_{12}\, \ln \alpha_{34}
\\&\hspace*{30pt}-2\ln \alpha_{12} \, U_1(\alpha_{23}) \, U_1(\alpha_{34})
-2\ln \alpha_{34}\, U_1(\alpha_{12}) \, U_1(\alpha_{23})
+4\ln \alpha_{23}\, U_1(\alpha_{12})\, U_1(\alpha_{34})\Bigg]\,,
\end{split}
\end{align}
while that of the 1-1-1-3 subtracted web of figure~\ref{1113} is
\begin{align}
\begin{split}
\overline{w}^{(3,-1)}_{(123444)}=-\frac16
\left(\frac{1}{4\pi}\right)^3\,
T_1^aT_2^bT_3^cT_4^d \bigg[
&f^{ade}f^{bce} F(\alpha_{14},\alpha_{24},\alpha_{34})
+f^{ace}f^{bde}
 \, F(\alpha_{24},\alpha_{14},\alpha_{34})
\bigg]
\end{split}
\end{align}
with
\begin{align}
\label{F1113_sym}
\begin{split}
F&(\alpha_{14},\alpha_{24},\alpha_{34})=
r(\alpha_{14})\,r(\alpha_{24})\,r(\alpha_{34})
\,\,\times \\&\bigg[
8 U_2(\alpha_{14})\, \ln \alpha_{24}\, \ln \alpha_{34}
+8 U_2(\alpha_{34}) \,\ln \alpha_{14}\, \ln \alpha_{24}
-16 U_2(\alpha_{24})\,\ln \alpha_{14}\, \ln \alpha_{34}
\\&+2\ln \alpha_{14} \, U_1(\alpha_{24}) \, U_1(\alpha_{34})
+2\ln \alpha_{34}\, U_1(\alpha_{14}) \, U_1(\alpha_{24})
-4\ln \alpha_{24}\, U_1(\alpha_{14})\, U_1(\alpha_{34})\bigg]\,.
\end{split}
\end{align}

\section{Conclusions}

In this talk we reviewed recent progress in the study of soft gluon exponentiation and long-distance singularities. The first part of the talk illustrated the way in which the non-Abelian exponentiation theorem gets generalised when considering a product of more than two Wilson lines. A first step was taken in Ref.~\cite{Gardi:2010rn} where an algorithm for computing exponentiated colour factors was derived using the replica trick. This led to formulating exponentiation in terms of webs which consist of sets of diagrams rather than individual ones, followed by the discovery~\cite{Gardi:2010rn,Mitov:2010rp,Gardi:2011wa,Gardi:2011yz,Dukes:2013wa,Dukes:2013gea,Gardi:2013ita} of several interesting properties of webs, which relate to their colour structure on the one hand and to their renormalization properties on the other. Recently it was shown in complete generality~\cite{Gardi:2013ita} that the colour factors in the exponent, for any web, are fully connected. 
The effective vertex formalism (see example in figure~\ref{Vn_connected}) developed in order to prove this theorem also led a natural basis for the colour factors, allowing to classify the connected colour factors which each web contributes to.

The second part of the talk was dedicated to the recent progress in evaluating the integrals of multi-gluon-exchange webs~\cite{Gardi:2013saa}.
We showed that organizing the calculation in terms of webs, and then subtracted webs, is essential in performing such integrals; not only do subtracted webs evaluate to substantially simpler expressions than individual diagrams, it is only at the level of the subtracted web that infrared-reguarization invariance is fully recovered and physical properties such as crossing symmetry are realised. 
We discussed the analytic structure of multi-gluon-exchange webs and illustrated how this structure was inferred. A major conjecture of Ref.~\cite{Gardi:2013saa} is that the symbol alphabet of all multi-gluon-exchange webs is $\left\{\ln \alpha_{ij},\, \ln(1-\alpha_{ij}^2)\right\}$. This in turn led to the conclusion that the result for any subtracted web in this class takes the form of a sum of products of polylogarithmic functions, each depending on a single cusp angle. The basis of functions was constructed for the class of four-line three-loop webs. Finally, explicit calculations of these webs were performed, confirming the expectations with regards to the analytic structure. 
The method of Ref.~\cite{Gardi:2013saa} is currently being used in the calculation of several other three- and four-loop webs~\cite{Falcioni:2013tb}. This, along with other recently developed techniques for dealing with multiple polylogarithms~\cite{Goncharov:arXiv0908.2238,Goncharov:2010jf,Duhr:2011zq,Duhr:2012fh}, which proves essential in the calculation of connected webs~\cite{Almelid:2013tb}, will eventually allow a complete calculation of the three-loop anomalous dimension, and lead to better understanding of long-distance singularities of scattering amplitudes.   

\vspace{20pt}
\noindent {\bf Acknowledgements:}
I would like to thank Samuel Abreu, \O yvind Almelid, Adrian Bodnarescu, Ruth Britto, Lance Dixon, Claude Duhr, Mark Dukes, Mark Harley, Johannes Henn, Gregory Korchemsky, Lorenzo Magnea, Mairi McKay, Jenni Smillie and Chris White, for stimulating discussions and collaboration on related projects. 
The research reported here is supported in part by the STFC grant ``Particle Physics at the Tait Institute''.

\bibliographystyle{JHEP}
\bibliography{refs3}

\end{document}